\documentclass[sigconf, nonacm]{acmart}
\usepackage{multirow}
\usepackage{enumerate}

\newcommand\vldbpagestyle{plain}
\usepackage{tikz}
\usepackage{bbding}
\usetikzlibrary{matrix,shapes,arrows,positioning,chains, calc,backgrounds}
\usetikzlibrary{decorations.pathreplacing}
\usepackage[misc]{ifsym}
\usepackage{subfigure}
\usepackage{cases}
\usepackage{amsmath}
\usepackage{multirow}
\usepackage{algorithm}
\usepackage{amsmath}

\begin{document}
\title{Quantum Identity-Based Encryption from the Learning with Errors Problem}
\author{Wenhua Gao$^{1,2,3}$, Li Yang$^{1,2,3}$, Daode Zhang, Xia Liu$^{1,2,3}$}
\affiliation{%
  \institution{$^1$State Key Laboratory of Information Security, Institute of Information Engineering, CAS, Beijing, China}
  \institution{$^2$School of Cyber Security, University of Chinese Academy of Sciences, Beijing, China}
  \institution{$^3$Institute of Information Engineering, Chinese Academy of Sciences, Beijing, China}
}
\email{gaowenhua@iie.ac.cn, yangli@iie.ac.cn}

\begin{abstract}
In order to prevent eavesdropping and tampering, the network security protocols use a handshake with an asymmetric cipher to establish a session-specific shared key with which further communication is encrypted using a symmetric cipher. The commonly used asymmetric algorithms include public key encryption, key exchange and identity-based encryption (IBE). However, the network security protocols based on classic identity-based encryption do not have perfect forward security. To solve the problem, we construct the first quantum IBE (QIBE) scheme based on the learning with errors problem, and prove that our scheme is fully secure under the random oracle. Moreover, we construct the quantum circuit of our QIBE scheme and give an estimate of the quantum resource of our circuit including the numbers of Hadamard gate, phase gate, T gate, CNOT gate and the total qubits used in the circuit, and conclude that the quantum resources required by our scheme increase linearly with the number of bits of the encrypted quantum plaintext. Our scheme exhibits the following advantages:\\
\indent$\bullet$ The classic key generation center (KGC) system still can be used for our QIBE scheme to generate and distribute the secret identity keys so that the cost can be reduced when the scheme is implemented. The reason why the classic KGC can be used is that the public and private keys are in the form of classic bits. \\
\indent$\bullet$ The network security protocols using a handshake with our QIBE scheme can provide perfect forward security. In our scheme, the ciphertext is transmitted in the form of a quantum state that is unknown to the adversary and therefore cannot be copied and stored. Thus, in the network security protocols based on our QIBE construction, the adversary cannot decrypt the previous quantum ciphertext to threat the previous session keys even if the identity secret key is threatened. \par

\end{abstract}

\maketitle

\pagestyle{\vldbpagestyle}
\begingroup
\renewcommand\thefootnote{}\footnote{\noindent
${(\textrm{\Letter})}$ Li Yang \\
~~~~yangli@iie.ac.cn\\
}\addtocounter{footnote}{-1}\endgroup

\section{Introduction}
\label{intro}
Identity-based cryptosystem is a public key cryptosystem first proposed by Shamir in 1984 \cite{DBLPShamir84}, whose public key is calculated directly from the receiver's identity $id_R$ such as phone number, email address, or network address, and the corresponding secret key $sk_R$ is calculated by the trusted key generation center (KGC) who owns the master public key $\mathsf{mpk}$ and master secret key $\mathsf{msk}$. When the sender wants to send message $m$ to the receiver, the sender encrypts the message to get the ciphertext $c=\mathsf{Encrypt}(\mathsf{mpk},id_R, m; r)$, where $r$ is a random number. On receiving the ciphertext $c$, the receiver can decrypt and get the message $m=\mathsf{Decrypt}(sk_R,c)$. Compared with cryptographic systems based on public key infrastructure (PKI), identity-based cryptosystems avoid the high cost of storing and managing public key certificates, simplify the management process of public keys, and reduce the pressure on the system. Therefore, identity-based cryptosystems have been widely developed and applied.\par
 The first practical identity-based encryption (IBE) scheme was proposed by Boneh et al. \cite{DBLPcryptoBonehF01} in 2001, which was followed by numerous other classic IBE schemes. These classic identity-based encryption (IBE) schemes can be mainly divided into three categoreies: IBE schemes based on elliptic curve bilinear mapping \cite{DBLPcryptoBonehF01,DBLP:conf/eurocrypt/BonehB04a,DBLP:conf/eurocrypt/Waters05}, IBE schemes based on quadratic residue \cite{DBLP:conf/ima/Cocks01,DBLP:conf/focs/BonehGH07,DBLP:conf/cisc/JhanwarB08,DBLP:conf/pkc/Joye16}, and IBE schemes based on lattices \cite{DBLP:conf/stoc/GentryPV08,DBLP:conf/eurocrypt/CashHKP10,DBLP:conf/eurocrypt/AgrawalBB10,DBLP:conf/scn/XieXZ12,DBLP:conf/eurocrypt/Yamada16}. With the development of quantum computers and quantum algorithms, especially the proposal of Shor algorithm \cite{DBLP:journals/siamcomp/Shor97}, the security of IBE schemes based on elliptic curve bilinear pair and quadratic residue have been seriously threatened. Since there is no quantum algorithm that can solve lattice-based difficult problems, the design and research of lattice-based IBE schemes have become the research hotspot of cryptographers.\par
 \vspace{2mm}

\tikzset{global scale/.style={
    scale=#1,
    every node/.append style={scale=#1}
  }
}

\begin{figure}[h]
	\begin{tikzpicture}[global scale =0.62,framed]
	\matrix(m)[matrix of nodes, column  sep=0cm,row  sep=0mm, nodes={draw=none, text depth=0pt},
	column 1/.style={anchor=base east},
	column 2/.style={anchor=base},
	column 3/.style={anchor=base west}]{
		\textbf{Sender}~                               &               &\textbf{Receiver}~\\
	                                                       & Hello~~~~~~        &\\
                                                           &                &\\
                                                           &               &\\
                                                           &              & ~~\\
		                                                   &  $(id_{R},\mathsf{mpk})$ &            \\
$sessionKey \overset{\$}{\leftarrow} \{0,1\}^{\ast}$      &                   &\\
$r\overset{\$}{\leftarrow} \{0,1\}^{\ast}$ &                                             &\\
$c=\mathsf{Encrypt}(\mathsf{mpk},id_{R},sessionKey;r)$          &                                             & \\
 ~~                    &                                             &\\
                                                           & $c$~~~~~ &\\
                                                           &     &  $sessionKey=\mathsf{Decrypt}(sk_{R},c)$ \\
                                                           &  Secure communication  & \\
                                                           &     &  \\
                                                           &     &  \\
                                                           &      & \\
                                                           &      &\\ 				
	};
	\draw[shorten <=-0cm,shorten >=-1cm] (m-1-1.south east)--(m-1-1.south west);
	\draw[shorten <=-0.7cm,shorten >=-0cm] (m-1-3.south east)--(m-1-3.south west);
    \draw[shorten <=-0.65cm, shorten >=-0.65cm,-latex',fill] (m-2-2.south west)--(m-2-2.south east);
	\draw[shorten <=-0.40cm,shorten >=-0.40cm,-latex',fill] (m-6-2.south east)--(m-6-2.south west);
	\draw[shorten <=-0.9cm,shorten >=-0.9cm,-latex',fill] (m-11-2.south west)--(m-11-2.south east);
    \draw [<->]  (m-13-2.south west)--(m-13-2.south east);
	\end{tikzpicture}
\caption{\small{Security protocols based on IBE}}
 \label{Security protocols}
 \centering
\end{figure}
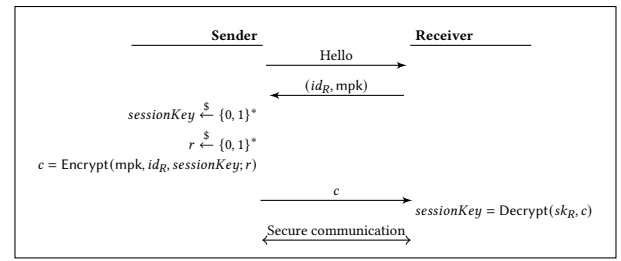

There are many applications of IBE such as constructing network security protocol (Chinese SSL VPN technology specification \cite{SSLvpn}). In the security protocol based on IBE, the receiver will send their identity $id_R$ and $\mathsf{mpk}$ to the sender, and the sender chooses a $sessionKey$ and sends its ciphertext to the receiver. Then the receiver decrypts the ciphertext to obtain the $sessionKey$. After that, both of them can own this secret key $sessionKey$ with which further communication is encrypted using a symmetric cipher. The above process is briefly described in Figure \ref{Security protocols}. A security protocol is said to provide perfect forward secrecy \cite{van1996handbook} if the compromise of long-term keys does not compromise past session keys that have been established before the compromise of the long-term key. In the security protocol based on classic IBE, all session keys and their ciphertexts are in the form of classic bits. A patient attacker can capture conversations to store the ciphertexts of session keys whose confidentiality is protected by the secret identity key (which is called the long-term key) and wait until the long-term key is threatened. Once the patient attacker gets the long-term key, they can decrypt the ciphertext of all previous session keys. In a word, all encrypted communications and sessions recorded in the past can be retrieved. Therefore, the security protocol based on classic IBE does not have perfect forward security. To solve this problem, considering that an adversary cannot replicate an unknown quantum state \cite{1982A}, we construct an quantum identity-based encryption (QIBE) scheme based on learning with errors problem. In our QIBE scheme, the ciphertext is transmitted in the form of a quantum state that is unknown to the adversary, and the ciphertext of session keys can not be copied. Then, in the security protocol based on our QIBE construction, even if the secret identity key is threatened, the adversary does not have the previous ciphertexts of session keys to decrypt so that they can not threat the security of the previous session keys. Therefore, the security protocol based on our QIBE construction has perfect forward security.\par
\subsection{Our Contributions}
In this work, we give the definition of identity-based quantum encryption and construct the first QIBE scheme based on the proposed classic identity-based encryption scheme \cite{DBLP:conf/stoc/GentryPV08}, and proved that our scheme is fully secure under the random oracle.\par
In our scheme, the $\mathsf{Setup}$ and $\mathsf{KeyGen}$ algorithms are classic algorithms and then the public and private keys are classic bits. Thus the classic key generation center (KGC) system still can be used for our QIBE scheme to generate and distribute the secret identity keys so that the cost can be reduced when the scheme is implemented. \par
In our scheme, the ciphertext is transmitted in the form of a quantum state that is unknown to the adversary and cannot be copied and stored. Therefore, in the network security protocols using a handshake with our QIBE scheme, if the identity private key is threatened, the adversary cannot decrypt the previous quantum ciphertext state to threat the previous session keys. Therefore, the network security protocol based on our QIBE can have perfect forward security compared to the network security protocol based on the classic IBE.\par
we construct the quantum circuit of our QIBE scheme and give an estimate of the quantum resource of our circuit including the numbers of Hadamard gate, phase gate, T gate, CNOT gate and the total qubits used in the circuit, and conclude that the quantum resources required by our scheme increase linearly with the number of bits of the encrypted quantum plaintext.\par
\par



\subsection{Outline of the paper}
The remainder of this paper is organised as follows: Section \ref{Preliminaries} describes the basic notation and previous work on quantum circuit, and basic knowledge and definitions of classic IBE and lattices. Section \ref{Quantum IBE} gives the definition of QIBE, describes the concrete construction of our scheme, analyzes the correctness of our scheme and gives its security proof, and analyzes the forward security of the network protocol based on our scheme. Section \ref{Quantum Circuit Realization of Quantum IBE} constructs the specific quantum circuit of our QIBE scheme, and estimates the quantum resources needed. Section \ref{Conclusion} summarises our work and presents directions for future work.
\section{Preliminaries}
\label{Preliminaries}
\begin{figure}[h]
 \includegraphics[width=0.4\textwidth]{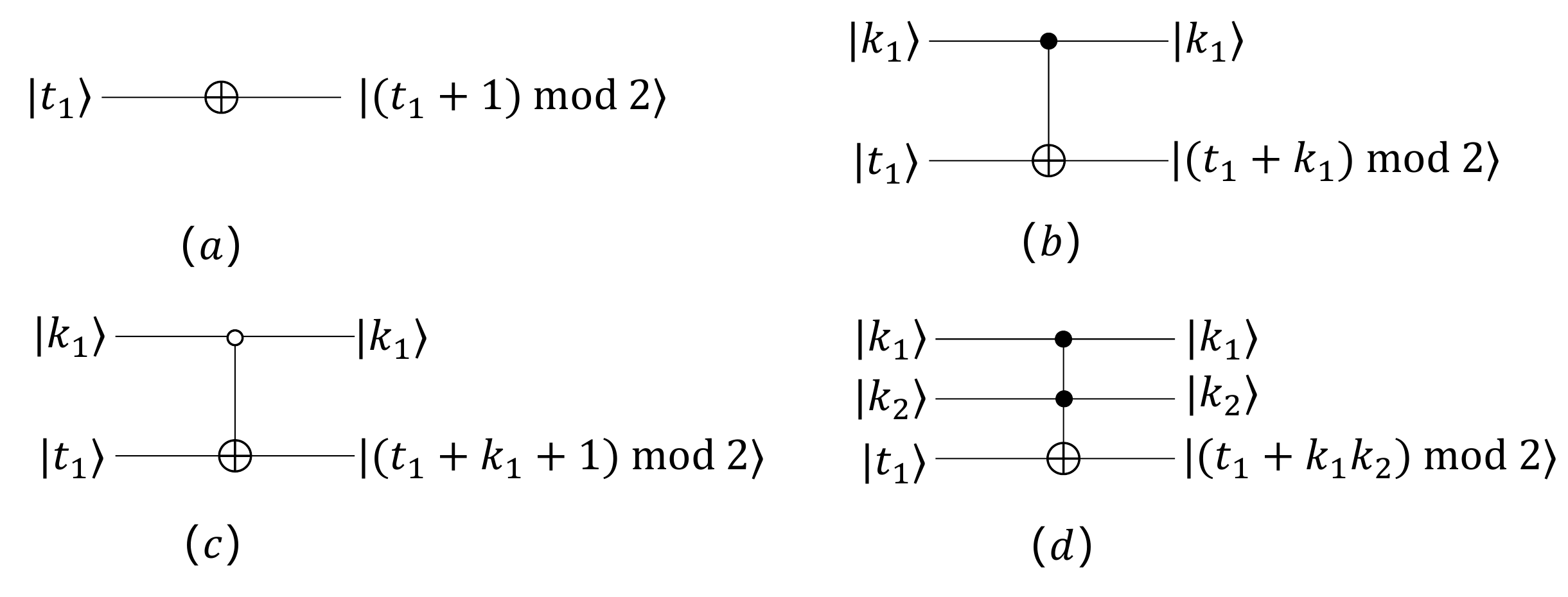}\\
 \caption{\small{The Basic quantum gates}}
\label{quantum gates}
 \centering
\end{figure}

The basic quantum gate involved in this study includes the single-qubit gate the NOT gate shown in Figure \ref{quantum gates}.($a$), double-qubit gates the CNOT gate shown in Figure \ref{quantum gates}.($b$) and a variant of it shown in \ref{quantum gates}.($c$) which can be obtained by a CNOT gate and two NOT gates, and three-qubit gate the Toffoli gate shown in Figure \ref{quantum gates}.($d$).

\begin{figure}[h]
 \includegraphics[width=0.3\textwidth]{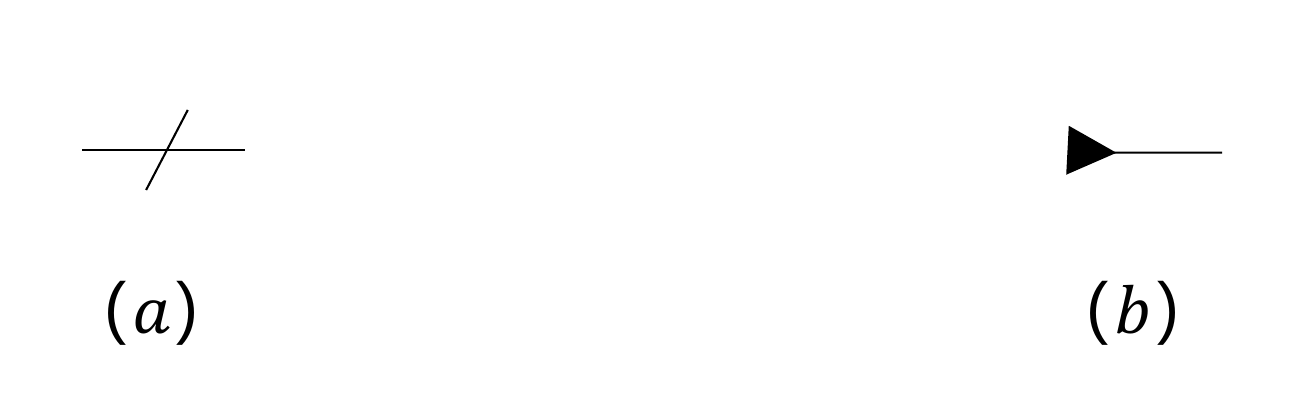}\\
 \caption{\small{The involved notations}}
\label{notations}
 \centering
\end{figure}

The notations involved in this study is shown in Figure \ref{notations}, where ($a$) means the input is multiple qubits and the black triangles in ($b$) represents the main output registers.\\
\\
We use bold lowercase letters to denote vectors. On the basis of the above, the $l$-controlled NOT gate and the combination of multi-CNOT gates involved in this study are shown in Figure 4, where ($a$) is the $l$-controlled NOT gate and its simplified form, which can be decomposed into $2l-3$ Toffoli gates, and ($b$) is a combination of $l$ CNOT gates and its simplified form. In addition, the simplified circuit of controlled copying the classical constant $d$  which can be decomposed into $l$ bit binary string $\mathbf{t}\in\{0,1\}^l$ is shown in Figure \ref{copying}. This circuit is implemented by performing CNOT operation or not according the value of each bit of $\mathbf{t}$ is one or zero. If $t_i=1(i=1,...,l)$, take $|k_1\rangle$ as control bit and the $i-$th bit of $|\mathbf{0}\rangle$ as target bit to perform CNOT operation; If $t_i=0$, do not any operation to the $i-$th bit of $|\mathbf{0}\rangle$. Finally, the output will produce $(|k_1\rangle,|d\ast k_1\rangle)$. In general, zero and one in $\mathbf{t}$ are approximately uniform, so this circuit requires approximately ${l}/{2}$ CNOT gates.\\

\begin{figure}[h]
 \includegraphics[width=0.48\textwidth]{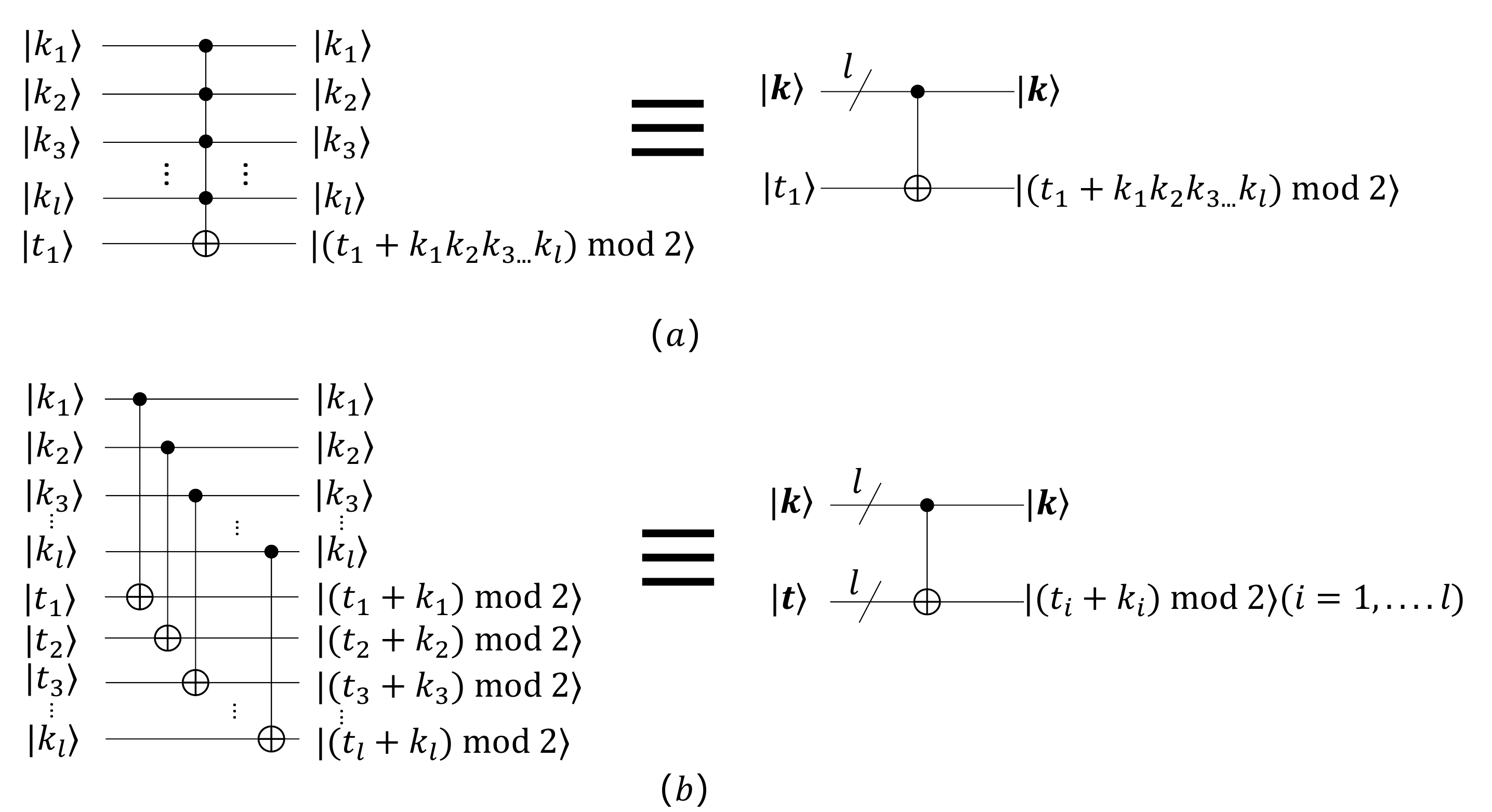}\\
 \caption{\small{$l$-controlled NOT gate and the combination of multi-CONT gates}}
\label{combinition}
 \centering
\end{figure}

\begin{figure}[h]
 \includegraphics[width=0.3\textwidth]{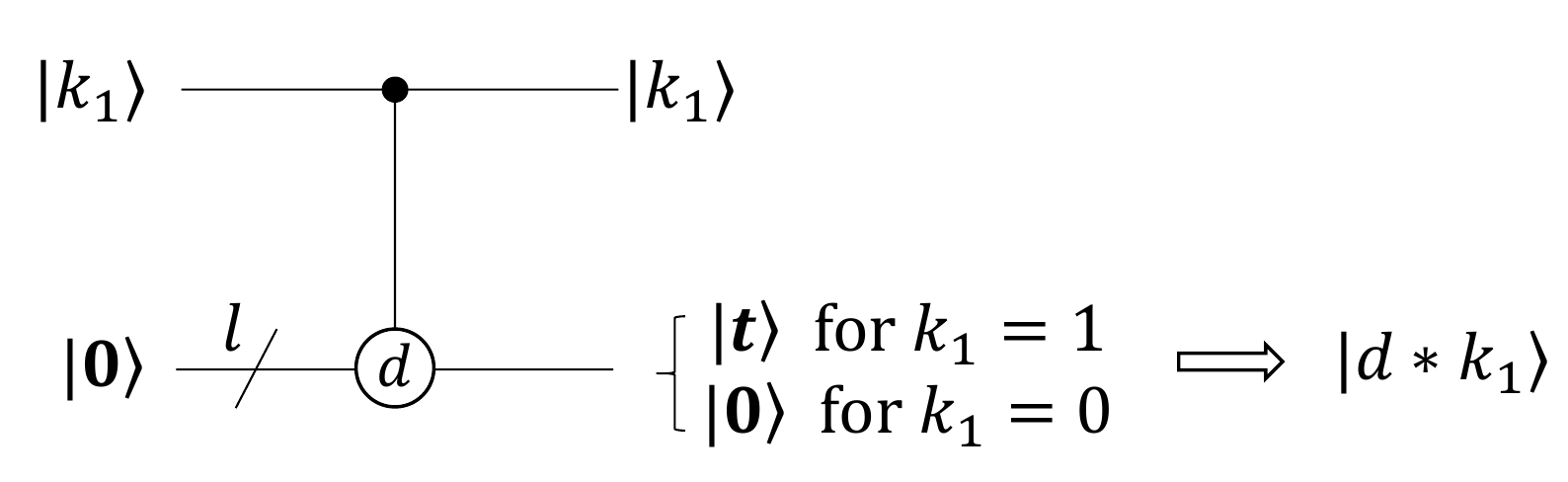}\\
 \caption{\small{controlled copying the classical constant $d$ circuit}}
\label{copying}
 \centering
\end{figure}
\subsection{Quantum Circuit}
\label{Quantum Circuit}
 The transformation of quantum states is realized by a series of unitary operations, which can be decomposed into many elementary gate operations. Therefore, the realization of quantum circuits is also accomplished by a series of gate operations.
 In this section, we describe the proposed quantum arithmatic operations including addition and subtraction, controlled addition, modular addition and comparison, and their corresponding quantum resources required including the numbers of CNOT gate, the Toffoli gate and the total qubit. All these works lay the foundation for the quantum circuit realization of our QIBE. To simplify the description, we show the simplified form of these arithmatic operations here and their specific implementation process can be seen in corresponding reference.\par

\begin{figure}[h]
 \includegraphics[width=0.42\textwidth]{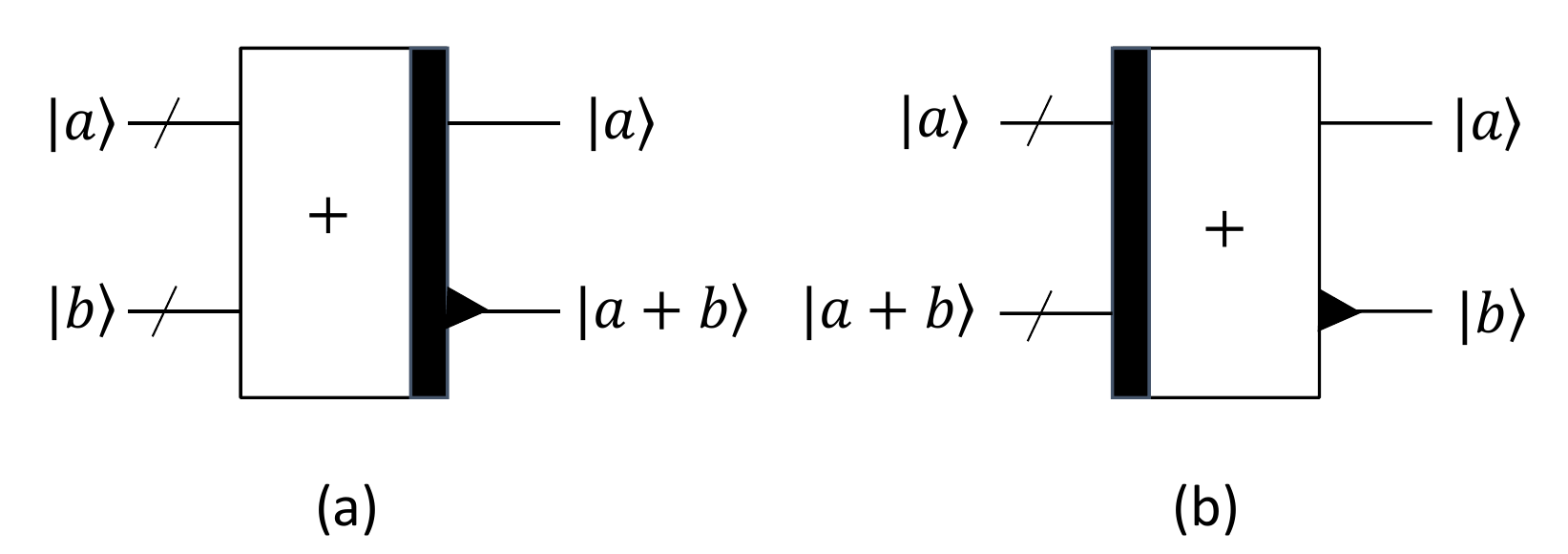}\\
 \caption{\small{simplified form of the quantum addition and subtraction network}}
\label{adder and subtractor}
 \centering
\end{figure}

\indent$\bullet$~\textbf{Addition and subtraction :}  Cuccaro et al. proposed a quantum addition circuit \cite{cuccaro2004new}. The quantum addition achieves the addition of two registers, that is $$|a,b\rangle\rightarrow |a,a+b\rangle.$$ To prevent overflows caused by carry, the second register (initially loaded in state $|b\rangle$) should be sufficiently large, i.e. if both $a$ and $b$ are encoded on $l$ qubits, the second register should be of size $l+1$. In the addition network, the last carry is the most significant bit of the result and is written in the $l+1$-th qubit of the second register. Because of the reversibility of unitary operations, by reversing the network of addition, i.e., apply each gate of the network in the reversed order, the subtraction network will be obtained.
The simplified form of the addition and subtraction network are shown in $(a)$ and $(b)$ of Figure \ref{adder and subtractor}. In this paper, a network with a bar on the left side represents the reversed sequence of elementary gates embedded in the same network with the bar on the right side.
On the subtraction network, with the input $(|a\rangle,|b\rangle)$, the output will produce $(|a\rangle,|a-b\rangle)$ when $a\geq b$. When $a< b$, the output is $(|2^{l}-(b-a)\rangle)$, where the size of the second register is $l+1$. i.e.,
 \begin{equation*}
\begin{cases}
|a,b\rangle\rightarrow |a,a-b\rangle,~~~~\mbox{for}~~a\geq b.\\
|a,b\rangle\rightarrow |a,2^{l}-(b-a)\rangle,~~~~\mbox{for}~~a< b.
\end{cases}
\end{equation*}
 When $a<b$, the significant qubit, the $l+1$-th qubit of the second register, which indicates whether or not an overflow occurred in the subtraction, will always contains 1. To calculate the addition or subtraction of two $l$-bit length inputs, a total of $2l$ Toffoli gates, $4l+1$ CNOT gates, and a total of $2l+2$-qubit are required for the addition or subtraction network.
 \par
\begin{figure}[h]
 \includegraphics[width=0.48\textwidth]{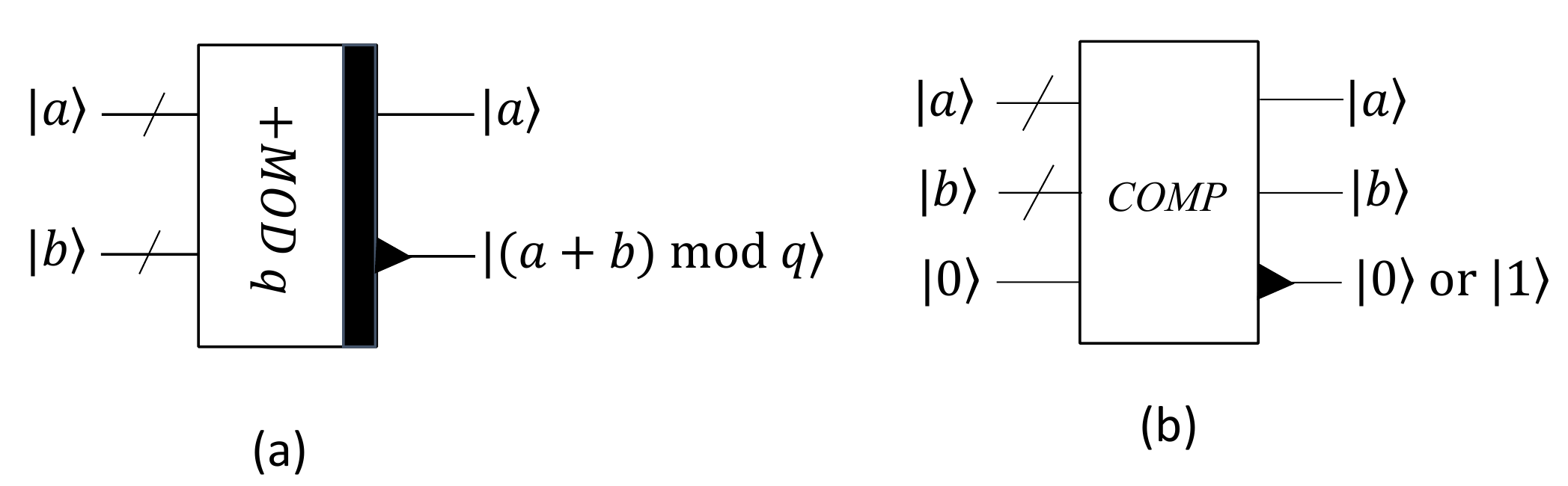}\\
 \caption{\small{simplified form of the quantum quantum adder modulo $q$ network and quantum comparison network}}
\label{comparison}
 \centering
\end{figure}

\indent$\bullet$~\textbf{Addition module $q$ :} Liu et al. \cite{liuxia20212} improved Roetteler's \cite{2017Quantum11} quantum modular addition circuit, reducing the number of quantum gates required. This quantum network effects
$$|a,b\rangle\rightarrow |a,(a+b)~\textrm{mod}~q \rangle, $$
where $0\leq a,b<q$. The simplified form of the addition module $q$ network are shown in $(a)$ of Figure \ref{comparison}. The modular subtraction can be obtained by reversing modular addition circuit and its bar is on the left hand. To calculate the addition or subtraction module $q$ of two $\lfloor\log q+1\rfloor$-bit length inputs, a total of $8\lfloor\log q+1\rfloor$ Toffoli gates, $13\lfloor\log q+1\rfloor+6$ CNOT gates and $3\lfloor\log q+1\rfloor+3$-qubit are required for this addition or subtraction module $q$ network.\par
 \indent$\bullet$~\textbf{Comparison :} Markov et al. construct a quantum comparison circuit by comparing $|a\rangle$ and $|b\rangle$ by whether the highest bit of $|a-b\rangle$ is $|0\rangle$ or $|1\rangle$ \cite{2012Constant}. This circuit is obtained by modifying the the previous subtraction circuit so that it outputs only the highest bit of $|a-b\rangle$. The comparison network achieves the comparison of two registers, that is \begin{equation*}
\begin{cases}
|a,b\rangle|0\rangle\rightarrow |a,b\rangle|0\rangle,~~~~\mbox{for}~~a\geq b.\\
|a,b\rangle|0\rangle\rightarrow |a,b\rangle|1\rangle,~~~~\mbox{for}~~a< b.
\end{cases}
\end{equation*}
The simplified form of the quantum comparison network is shown in Figure \ref{comparison}.$(b)$. To comparing two $l$-bit length inputs $|a\rangle$ and $|b\rangle$, a total of $2l$ Toffoli gates, $4l+1$ CNOT gates, and $2l+2$ qubits are required for the comparison network.\par

\subsection{Lattices}
\label{Lattices}
 Let $X$ and $Y$ be two random variables over some finite set $S_X$, $S_Y$, respectively. The statistical distance $\Delta(X,Y)$ between $X$ and $Y$ is defined as $$\Delta(X,Y)=\frac{1}{2}\sum_{s\in {S_X\cup S_Y}}\left|\Pr[X=s]-\Pr[Y=s]\right|.$$\par
 For integer $q\geq2$, $\mathbb{Z}_q$ denotes the quotient ring of integer modulo $q$. We use bold capital letters to denote matrices, such as $\mathbf{A,B}$, and bold lowercase letters to denote vectors, such as $\mathbf{x,y}$. We denote the $j$-th row of a matrix $\mathbf{R} $ by $\mathbf{r}_j$ and its $i$-th column by $\mathbf{r}^i$. Moreover, we denote the $j$-th element of a vector $\mathbf{m} $ by $m_j$. The notations $\mathbf{A}^\top$ denote the transpose of the matrix $\mathbf{A}$. Specially, we use $\mathbf{i}$ to denote a vector that each element is one, i.e, $\mathbf{i}=(1,\cdots, 1)^\top$.\par
 Let $\mathbf{S}$ be a set of vectors $\mathbf{S}=\{\mathbf{s}_{1},\cdots,\mathbf{s}_{n}\}$ in $\mathbb{R}^{m}$. We use $\widetilde{\mathbf{S}}=\{\widetilde{\mathbf{s}}_{1},\cdots,\widetilde{\mathbf{s}}_{n}\}$ to denote the Gram-Schmidt orthogonalization of the vectors $\mathbf{s}_{1},\cdots,\mathbf{s}_{n}$ in that order, and $\|\mathbf{S}\|$ to denote the length of the longest vector in $\mathbf{S}$. For positive integers $q,n,m$ with $q$ prime, and a matrix $\mathbf{A}\in\mathbb{Z}_{q}^{n\times m}$, the $m$-dimensional integer lattices are defined as: $\Lambda_{q}(\mathbf{A})=\{\mathbf{y}:\mathbf{y}=\mathbf{A}^{\top}\mathbf{s}~\mathrm{for}~\mathrm{some}~\mathbf{s}\in\mathbb{Z}^{n}\}$ and $\Lambda_{q}^{\perp}(\mathbf{A})=\{\mathbf{y}:\mathbf{A}\mathbf{y}=\mathbf{0}\mod q\}$. Moreover, for $\mathbf{u}\in \mathbb{Z}_{q}^{n}$, the set of syndromes is defined as $\Lambda_{q}^{\mathbf{u}}(\mathbf{A})=\{\mathbf{y}:\mathbf{u}=\mathbf{A}\mathbf{y}\mod q\}$. \par

 For $\mathbf{x}\in\Lambda$, define the Gaussian function $\rho_{s,\mathbf{c}}(\mathbf{x})$ over $\Lambda\subseteq\mathbb{Z}^m$ centered at $\mathbf{c}\in\mathbb{R}^{m}$ with parameter $s>0$ as $\rho_{s,\mathbf{c}}(\mathbf{x})=\exp(-\pi||\mathbf{x-c}||/s^2)$. Let $\rho_{s,\mathbf{c}}(\Lambda)=\sum_{\mathbf{x}\in\Lambda}\rho_{s,\mathbf{c}}(\mathbf{x})$, and define the discrete Gaussian distribution over $\Lambda$ as $\mathcal{D}_{\Lambda,s,\mathbf{c}}(\mathbf{x})=\frac{\rho_{s,\mathbf{c}}(\mathbf{x})}{\rho_{s,\mathbf{c}}(\Lambda)}$, where $\mathbf{x}\in\Lambda$. For simplicity, $\rho_{s,\mathbf{0}}$ and $\mathcal{D}_{\Lambda,s,\mathbf{0}}$ are abbreviated as $\rho_{s}$ and $\mathcal{D}_{\Lambda,s}$, respectively.\par

\begin{lemma}
\label{SixLemma}
Let $q,n,m$ be positive integers with $q\geq2$ and $q$ prime. There exists PPT algorithms such that
\begin{itemize}
\item[$\bullet$] (\cite{Ajtai99,AlwenP09}): $\mathsf{TrapGen}(1^n,1^{m},q)$ a randomized algorithm that, when $m\geq6n\lceil\log q\rceil$, outputs a pair $(\mathbf{A,T_{A}})\in\mathbb{Z}_{q}^{n\times m}\times\mathbb{Z}^{m\times m}$ such that $\mathbf{A}$ is $2^{-\Omega(n)}-$close to uniform in $\mathbb{Z}_{q}^{n\times m}$ and $\mathbf{T_{A}}$ is a basis of $\Lambda^{\perp}_{q}(\mathbf{A})$, satisfying $\|\widetilde{\mathbf{T_{A}}}\|\leq\mathcal{O}(\sqrt{n\log q})$ with overwhelming probability.
\item[$\bullet$] (\cite{DBLP:conf/stoc/GentryPV08}): $\mathsf{SampleD}(\mathbf{A},\mathbf{T_{A}},\mathbf{u},\sigma)$  a randomized algorithm that, given a full rank matrix $\mathbf{A}\in \mathbb{Z}_{q}^{n\times m}$,a basis $\mathbf{T_{A}}$ of $\Lambda^{\perp}_{q}(\mathbf{A})$, a vector $\mathbf{u}\in \mathbb{Z}_{q}^{n}$ and $\sigma\geq\|\widetilde{\mathbf{T_{A}}}\|\cdot\omega(\sqrt{\log m})$, then outputs a vector $\mathbf{r}\in\mathbb{Z}_{q}^{m}$ sampled from a distribution $2^{-\Omega(n)}-$close to $\mathcal{D}_{\Lambda_{q}^{\mathbf{u}}(\mathbf{A}),\sigma}$.
\end{itemize}
\end{lemma}
\noindent $\mathbf{Discrete~Gaussian~Lemmas.}$ The following lemmas are used to manipulate and obtain meaningful bounds on discrete Gaussian vectors.

\begin{lemma}
\label{fourLemma}
(Adopted from \cite{DBLP:conf/stoc/GentryPV08}, Lem.5.2). Let $n$, $m$, $q$ be positive integers such that $m\geq 2n\log q$ and $q$ a prime. Let $\sigma$ be any positive real such that $\sigma\geq\sqrt{n+\log m}$. Then for all but $2^{-\Omega(n)}$ fraction of $\mathbf{A}\in\mathbb{Z}_q^{n\times m}$, we have that the distribution of $\mathbf{u}=\mathbf{Ar}~\textrm{mod}~q$ for $\mathbf{r}\leftarrow D_{\mathbb{Z}^m,\sigma}$ is $2^{-\Omega(n)}$-close to uniform distribution over $\mathbb{Z}_q^n$. Furthermore, for a fixed $\mathbf{u}\in \mathbb{Z}_q^n$, the conditional distribution of $\mathbf{r}\leftarrow D_{\mathbb{Z}^m,\sigma}$, given $\mathbf{Ar}=\mathbf{u}~\textrm{mod}~q$ is $\mathcal{D}_{\Lambda_{q}^{\mathbf{u}}(\mathbf{A}),\sigma}$.

\end{lemma}

The security of our construction is based on the learning with errors (LWE) hardness assumption. The LWE problem is a hard problem based on lattices defined by Regev \cite{regev2005lattices}, which is stated below: given an input $(\mathbf{A},\mathbf{d})$, where $\mathbf{A}\in \mathbb{Z}_q^{n\times m}$ for any $m=poly(n)$ and integer $q\geq 2$ is prime and $\mathbf{d}\in \mathbb{Z}_q^m$ is either of the form $\mathbf{d}= (\mathbf{A}^\top \mathbf{s}+\mathbf{e})~\textrm{mod}~q$ for $\mathbf{s}\in \mathbb{Z}_q^n$ and $\mathbf{e}\in \mathcal{D}_{\mathbb{Z}^m,\sigma}$ or is uniformly random (and independent of $\mathbf{A}$), distinguish which is the case, with non-negligible advantage. Regev proved that LWE problem is as hard as approximating standard lattice problems in the worst case using a quantum algorithm.

\par

\subsection{Classic Identity-Based Encryption}
\label{Identity-Based Encryption}
A classic IBE scheme consists of the following four algorithms:
\begin{itemize}
 \item[$\bullet$] $\mathsf{KeyGen}(1^\lambda)\rightarrow (\mathsf{mpk}, \mathsf{msk})$. The key generation algorithm takes in a security parameter $1^\lambda$ as input. It outputs master public key $\mathsf{mpk}$ and a master secret key $\mathsf{msk}$.
\item[$\bullet$] $\mathsf{Extract}(\mathsf{mpk}, \mathsf{msk}, id)\rightarrow sk_{id}$. The key extraction algorithm takes master public key $\mathsf{mpk}$, master secret key $\mathsf{msk}$, and identity ${id}$ as input. It outputs $sk_{id}$ as the secret key.
\item[$\bullet$] $\mathsf{Encrypt}(\mathsf{mpk},id,\mathsf{M};r)\rightarrow c$. The encryption algorithm takes in public parameters $\mathsf{mpk}$, identities $id$, and a message $\mathsf{M}$ as input. It outputs a ciphertext $c$.
\item[$\bullet$] $\mathsf{Decrypt}(sk_{id},c) \rightarrow \mathsf{M}$. The decryption algorithm takes in the secret key $sk_{id}$ and a ciphertext $c$, as input. It outputs a message $\mathsf{M}$.
\end{itemize}
\emph{Correctness.} For all $(\mathsf{mpk},\mathsf{msk})\overset{\$}{\leftarrow} \mathsf{KeyGen}(1^\lambda)$, all identities $id\in{ID}$, all messages $M$, all $c \leftarrow\mathsf{Encrypt}(\mathsf{mpk},id,\mathsf{M};r)$, we have
$$
\Pr[\mathsf{Decrypt}(\mathsf{mpk},sk_{id},c)=\mathsf{M}]=1-\mathsf{negl}(\lambda).$$
\emph{Security.} The security game is defined by the following experiment, played by a challenger and an adversary $\mathcal{A}$:
\begin{enumerate}
\item The challenger runs $\mathsf{KeyGen}$ to generate $(\mathsf{mpk}, \mathsf{msk})$. It gives $\mathsf{mpk}$ to the adversary $\mathcal{A}$.
\item The adversary $\mathcal{A}$ adaptively requests keys for any identity $id_i$ of its choice. The challenger responds with the corresponding secret key $sk_{id_i}$, which it generates by running $\mathsf{Extract}(\mathsf{mpk}, \mathsf{msk}, id_i)$.
\item The adversary $\mathcal{A}$ submits two messages $\mathsf{M_0}$ and $\mathsf{M_1}$ of equal length and a challenge identity $id^\ast$ with the restriction that $id^\ast$ is not equal to any identity requested in the previous phase. The challenger picks $\beta \overset{\$}{\leftarrow} \{0,1\}$, and encrypts $\mathsf{M_\beta}$ under $id^\ast$ by running the encryption algorithm. It sends the ciphertext to the adversary $\mathcal{A}$.
\item $\mathcal{A}$ continues to issue key queries for any identity $id_i$ as in step (2) with the restriction that $id_i \neq id^\ast$.
\item The adversary $\mathcal{A}$ outputs a guess $\beta^\prime$ for $\beta$.
\end{enumerate}
\noindent The advantage $\mathsf{Adv}^{\mathrm{IBE}}_\mathcal{A}(\lambda)$ of an adversary $\mathcal{A}$ is defined to be $$\mathsf{Adv}^{\mathrm{IBE}}_\mathcal{A}(\lambda)=\left|\Pr [\beta^\prime=\beta]-1/2\right|.$$
$Definition~1$. An IBE scheme is fully secure if for all probabilistic polynomial-time adversaries $\mathcal{A}$, $\mathsf{Adv}^{\mathrm{IBE}}_\mathcal{A}(\lambda)$ is a negligible function in $\lambda$.

\section{Quantum IBE}
\label{Quantum IBE}
\subsection{Definition of QIBE}
\label{Definition of QIBE}
In this section, we give the definition of QIBE and classify QIBE.\par
{\bf Definition 2:}
If one or more elements of the quadruple $(\mathsf{KeyGen}, \mathsf{Extract}, \mathsf{Encrypt},\mathsf{Decrypt})$ of the IBE scheme are quantum process, then we call the IBE scheme quantum IBE, namely QIBE scheme.\par
It is analogous to the analysis and classification of quantum public key encryption \cite{wu2015complete} and quantum symmetric-encryption scheme \cite{xiang2014classification}, each element of the quadruple can be classic or quantum, so there may be sixteen types of QIBE schemes.

\subsection{Our Construction}
\label{Construction1}
In this section, we utilise the proposed classic IBE scheme \cite{DBLP:conf/stoc/GentryPV08} to construct a kind of QIBE scheme based on quantum trapdoor one-way transformation \cite{yang2010quantum}, whose directly encrypted message is a multi-bit quantum state. In our QIBE scheme, the algorithms $\mathsf{KeyGen}$ and $\mathsf{Extract}$ are classic process and the algorithms $\mathsf{Encrypt}$ and $\mathsf{Decrypt}$ are quantum process. To make it easier to distinguish between classic IBE and QIBE, we denote that our scheme consists of four algorithms $\mathsf{QIBE}=(\mathsf{QKeyGen},\mathsf{QExtract}, \mathsf{QEncrypt},\mathsf{QDecrypt})$. In the scheme, let integer parameters $n=\mathcal{O}(\lambda), m=\mathcal{O}(n),\sigma=\mathcal{O}(n^{0.5}), q=\mathcal{O}(m^{3.5})$ according to \cite{DBLP:conf/stoc/GentryPV08}, where $\lambda$ is a security parameter. \par
\noindent$\bullet\mathsf{QKeyGen}:$ (1) Use the algorithm $\mathsf{TrapGen}(m,n,q)$ to select a uniformly random $n\times m-$matric $\mathbf{A}\in \mathbb{Z}_q^{n\times m}$ and $\mathbf{T_A}\in \mathbb{Z}_q^{m\times m}$ which is a good basis for $\Lambda_q^\perp(\mathbf{A})$. (2) Select a hash function $\mathsf{H} : \{0, 1\}^n \rightarrow \mathbb{Z}_q^{n\times n}$  which map an identity to an $n\times n-$matric. (3) Output the master key $\mathsf{mpk}=(\mathbf{A},q,m,n,\mathsf{H})$ and $\mathsf{msk}=(\mathbf{T_A})$. (4) In a word, $\mathsf{QKeyGen}(\lambda, q,m,n)\rightarrow (\mathsf{mpk}=(\mathbf{A},q,m,n,\mathsf{H}),\mathsf{msk}=(\mathbf{T_A}))$. \par
\noindent$\bullet\mathsf{QExtract}$: (1) Input $\mathsf{mpk}$, $\mathsf{msk}$ and an identity $id \in\{0,1\}^n$. (2) Compute $\mathbf{U}=\mathsf{H}(id)$ and use the algorithm $\mathsf{SampleD}$ to generate $sk_{id}=\mathbf{R}$ such that $\mathbf{r}^i=\mathsf{SampleD}(\mathbf{A},\mathbf{T_A}, \mathbf{u}^i,\sigma)$ for $i=1,\cdots,n$. It is clear $\mathbf{U}=\mathbf{A}\mathbf{R}\bmod q$. (3) In a word, $\mathsf{QExtract}(\mathsf{msk},\mathsf{mpk},id)\rightarrow sk_{id}=\mathbf{R}$.\par

\noindent$\bullet\mathsf{QEncrypt}$: (1) To encrypt an $n-$qubit quantum superposition state $\sum_\mathbf{m}\alpha_\mathbf{m}|\mathbf{m}\rangle$, input an identity $id$, $\mathsf{mpk}$ and the quantum message $\sum_\mathbf{m}\alpha_\mathbf{m}|\mathbf{m}\rangle$, where $|\mathbf{m}\rangle$ is the basis state of the quantum message length $n-$qubit. (2) Compute $\mathbf{U}=\mathsf{H}(id)$. (3) Choose a uniformly random $\mathbf{s}\leftarrow \mathbb{Z}_q^n$, $\mathbf{e}_0\in \mathcal{D}_{\mathbb{Z}^n,\sigma}$ and $\mathbf{e}\in \mathcal{D}_{\mathbb{Z}^m,\sigma}$. (4) Set $\mathbf{x}=(\mathbf{U}^\top \mathbf{s}+\mathbf{e_0})~\textrm{mod}~q$ and $\mathbf{c_1}=(\mathbf{A}^\top \mathbf{s}+\mathbf{e})~\textrm{mod}~q$. Then more processes are performed as follows:\\
$\star \mathrm{Step}~1$: Take each bit of quantum state $\sum_\mathbf{m}\alpha_\mathbf{m}|\mathbf{m}\rangle$ as the control bit and $|\mathbf{0}\rangle$ as the input, by the controlled copying classical constant $\lfloor \frac{q}{2}\rfloor$ circuit we can get
$$\sum_\mathbf{m}\alpha_\mathbf{m}\left|\mathbf{m}\right\rangle\left|\lfloor \frac{q}{2}\rfloor\mathbf{m}\right\rangle.$$
$\star \mathrm{Step}~2$: Take above result and $\mathbf{x}$ as quantum adder modulo $q$ network's inputs, we can get
$$\sum_\mathbf{m}\alpha_\mathbf{m}\left|\mathbf{m}\right\rangle\left|(\mathbf{x}+\lfloor \frac{q}{2}\rfloor\mathbf{m})~\textrm{mod}~q\right\rangle.$$
$\star \mathrm{Step}~3$:Unentangle the two registers of the above result to get
$$ |\psi\rangle=\sum_\mathbf{m}\alpha_\mathbf{m}\left|(\mathbf{x}+\lfloor \frac{q}{2}\rfloor\mathbf{m})~\textrm{mod}~q\right\rangle,$$ and the specific unentanglement process will be described in detail in Section \ref{Quantum Circuit Realization of Quantum IBE}.\\
(5) In a word, $\mathsf{QEncrypt}(id,\sum_\mathbf{m}\alpha_\mathbf{m}|\mathbf{m}\rangle)\rightarrow (c=(\mathbf{c_1},|\psi\rangle))$.\par

\noindent$\bullet\mathsf{QDecrypt}$: (1) Input the master public key $\mathsf{mpk}$, the private key $\mathbf{R}$, and the ciphertext $(\mathbf{c_1},|\psi\rangle)$. (2) Set $\mathbf{R}^\top\mathbf{c_1}~\textrm{mod}~q=\mathbf{y}\in \mathbb{Z}_q^n$. Then more processes are performed as follows:\\
$\star \mathrm{Step}~1$: Take $|\mathbf{y}\rangle$ and $ |\psi\rangle=\sum_\mathbf{m}\alpha_\mathbf{m}\left|(\mathbf{x}+\lfloor \frac{q}{2}\rfloor\mathbf{m})~\textrm{mod}~q\right\rangle$ as the inputs of the inverse of quantum adder modulo $q$ network, we can get
$$\sum_\mathbf{m}\alpha_\mathbf{m} \left|\left(\left(\mathbf{x}+\lfloor \frac{q}{2}\rfloor\mathbf{m}\right)~\textrm{mod}~q-\mathbf{y}\right)~\textrm{mod}~q\right\rangle.$$
$\star \mathrm{Step}~2$: Take $\lfloor\frac{q}{2}\rfloor$ and above result as the inputs of quantum subtraction network, we can get
$$\sum_\mathbf{m}\alpha_\mathbf{m} \left| \left( \left(\mathbf{x}+\lfloor \frac{q}{2}\rfloor\mathbf{m}\right)~\textrm{mod}~q-\mathbf{y}\right)~\textrm{mod}~q-\lfloor\frac{q}{2}\rfloor\cdot\mathbf{i} \right\rangle.$$
$\star \mathrm{Step}~3$: Take above result as the input of quantum absolute value circuit which will be described in section \ref{Quantum Circuit1}, we can get $$\sum_\mathbf{m}\alpha_\mathbf{m} \left| \mathsf{abs}\left(\left( \left(\mathbf{x}+\lfloor \frac{q}{2}\rfloor\mathbf{m}\right)\textrm{mod}~q-\mathbf{y}\right)~\textrm{mod}~q-\lfloor\frac{q}{2}\rfloor\cdot\mathbf{i} \right) \right\rangle.$$
$\star \mathrm{Step}~4$: Take above result, $\lfloor\frac{q}{4}\rfloor$ and $|0\rangle$ as the inputs of quantum comparison network, we can get $$\sum_\mathbf{m}\alpha_\mathbf{m} \left| \mathsf{abs}\left(\left( \left(\lfloor\mathbf{x}+ \frac{q}{2}\rfloor\mathbf{m}\right)\textrm{mod}~q-\mathbf{y}\right)~\textrm{mod}~q-\lfloor\frac{q}{2} \rfloor \cdot\mathbf{i}\right) \right\rangle\left|\mathbf{m}\right\rangle.$$
Next, we will unentangle the first and the second register of this quantum state.\\
$\star \mathrm{Step}~5$: Take the first register of above result as the input of the inverse of quantum absolute value circuit, we can get $$\sum_\mathbf{m}\alpha_\mathbf{m} \left| \left( \left(\lfloor\mathbf{x}+ \frac{q}{2}\rfloor\mathbf{m}\right)~\textrm{mod}~q-\mathbf{y}\right)~\textrm{mod}~q-\lfloor\frac{q}{2}\rfloor\cdot\mathbf{i}  \right\rangle\left|\mathbf{m}\right\rangle.$$
$\star \mathrm{Step}~6$: Take $\lfloor\frac{q}{2}\rfloor$ and the first register of above result as quantum addition network's inputs, we can get
$$\sum_\mathbf{m}\alpha_\mathbf{m} \left| \left( \left(\mathbf{x}+\lfloor \frac{q}{2}\rfloor\mathbf{m}\right)~\textrm{mod}~q-\mathbf{y}\right)~\textrm{mod}~q \right\rangle\left|\mathbf{m}\right\rangle.$$
$\star \mathrm{Step}~7$: Take $|\mathbf{y}\rangle$ and the first register of above result as the quantum adder modulo $q$ network's inputs, we can get
$$\sum_\mathbf{m}\alpha_\mathbf{m} \left| \left(\mathbf{x}+\lfloor \frac{q}{2}\rfloor\mathbf{m}\right)~\textrm{mod}~q \right\rangle\left|\mathbf{m}\right\rangle.$$
$\star \mathrm{Step}~8$: Take each bit of the second register of above result as the control bit and $|\mathbf{0}\rangle$ as the input, by the controlled copying classic constant $\lfloor\frac{q}{2}\rfloor$ circuit, we can get
$$\sum_\mathbf{m}\alpha_\mathbf{m} \left| \left(\mathbf{x}+\lfloor \frac{q}{2}\rfloor\mathbf{m}\right)~\textrm{mod}~q \right\rangle\left|\mathbf{m}\right\rangle\left|\lfloor\frac{q}{2}\rfloor\mathbf{m}\right\rangle.$$
$\star \mathrm{Step}~9$: Take the first register and the third register of above result as the inputs of the inverse of quantum adder modulo $q$ network, we can get
$$\sum_\mathbf{m}\alpha_\mathbf{m} \left|\mathbf{x}\right\rangle\left|\mathbf{m}\right\rangle\left|\lfloor\frac{q}{2}\rfloor\mathbf{m}\right\rangle.$$
Then, we need to unentangle the second and the third register of this result.\\
$\star \mathrm{Step}~10$: Take each bit of the second register of above result as the control bit, and the third register of above result as the input of controlled copying classic constant $\lfloor\frac{q}{2}\rfloor$ circuit, that's, by performing the inverse operation of step 8, we can get
$$\sum_\mathbf{m}\alpha_\mathbf{m} \left|\mathbf{m}\right\rangle\left|\mathbf{0}\right\rangle.$$
Then, quantum state $\sum_\mathbf{m}\alpha_\mathbf{m}\left|\mathbf{m}\right\rangle$ is no longer entangled with other registers and the decryption process is complete. (3) In a word, $\mathsf{QDecrypt}(id, \mathsf{mpk},\mathbf{R},(\mathbf{c_1},|\psi\rangle))\rightarrow \sum_\mathbf{m}\alpha_\mathbf{m}\left|\mathbf{m}\right\rangle$.\par

\subsection{Correctness}
Consider a ciphertext
$$(\mathbf{c_1},|\psi\rangle)=\left((\mathbf{A}^\top \mathbf{s}+\mathbf{e})~\textrm{mod}~q, \sum_\mathbf{m}\alpha_\mathbf{m}\left|(\mathbf{x}+\lfloor \frac{q}{2}\rfloor\mathbf{m})~\textrm{mod}~q\right\rangle\right)$$ of an $n$-qubit quantum superposition state $\sum_\mathbf{m}\alpha_\mathbf{m}|\mathbf{m}\rangle$, it is easy to see that $|\psi\rangle=\sum_\mathbf{m}\alpha_\mathbf{m}\left|\mathbf{c}_0\right\rangle$, where $\mathbf{c}_0=(\mathbf{x}+\lfloor \frac{q}{2}\rfloor\mathbf{m})~\textrm{mod}~q=\mathbf{U}^\top \mathbf{s}+\mathbf{e_0}+\lfloor \frac{q}{2}\rfloor\cdot\mathbf{m} ~\textrm{mod}~q$. In the step $2$ of $\mathsf{QDecrypt}$, it is clear that \begin{eqnarray*}
\begin{aligned}
&\left(\left(\mathbf{x}+\lfloor \frac{q}{2}\rfloor\mathbf{m}\right)~\textrm{mod}~q-\mathbf{y}\right)~\textrm{mod}~q-\lfloor\frac{q}{2}\rfloor\cdot\mathbf{i}\\
=&\left(\mathbf{c_0}-\mathbf{y}\right)~\textrm{mod}~q-\lfloor\frac{q}{2}\rfloor\cdot\mathbf{i}\\
\end{aligned}
\end{eqnarray*}
which equals to $\mathbf{b}$ in the $\mathsf{Decrypt}$ of Theorem 1. Then in the step $3$ of $\mathsf{QDecrypt}$, we compute the absolute value $\mathsf{abs}(\mathbf{b})$ of $\mathbf{b}$. Finally in the step $4$ of $\mathsf{QDecrypt}$, we compare $\mathsf{abs}(\mathbf{b})$ with $\lfloor\frac{q}{4}\rfloor$ and get $\mathbf{m}$. According to Theorem $1$, the decryption algorithm $\mathsf{Decrypt}$ with the identity secret key $sk_{id}=\mathbf{R}_{id}$ can decrypt the ciphertext $c=(\mathbf{c}_0,\mathbf{c}_1)$ correctly with a probability $1-\mathsf{negl}(\lambda)$. Therefore, the decryption algorithm $\mathsf{QDecrypt}$ with the identity secret key $sk_{id}=\mathbf{R}_{id}$ can decrypt the ciphertext $c=(\mathbf{c}_1,|\psi\rangle)$ correctly with a probability $1-\mathsf{negl}(\lambda)$.

\subsection{Security proof}
\label{Security analysis}
\begin{theorem}
\label{theorem2222}
The above IBE scheme $\mathrm{QIBE}$ is fully secure in the random oracle model assuming the hardness of $\mathrm{LWE}$. Namely, for any classical PPT adversary $\mathcal{A}$ making at most $Q_\mathsf{H}$ random oracle queries to $\mathsf{H}$ and $Q_{\mathrm{ID}}$ secret key queries, there exists a classical PPT algorithm $\mathcal{B}$ such that
\begin{equation}
\label{finalEquation}
 \mathrm{Adv}_{\mathcal{A}}^{\mathrm{QIBE}}(\lambda)\leq Q_{\mathsf{H}}\cdot\mathrm{Adv}_{\mathcal{B}}^{\mathrm{LWE}}(\lambda)+(n\cdot Q_\mathsf{H}+n\cdot Q_{\mathsf{ID}}+1)\cdot2^{-\Omega(n)}.
\end{equation}
\end{theorem}
\noindent\textbf{Proof} (of Theorem \ref{theorem2222}.) Without loss of generality, we make some simplifying assumptions on $\mathcal{A}$. First, we assume that whenever $\mathcal{A}$ queries a secret key or asks for a challenge ciphertext, the corresponding $id$ has already been queried to the random oracle $\mathsf{H}$. Second, we assume that $\mathcal{A}$ makes the same query for the same random oracle at most once. Third, we assume that $\mathcal{A}$ does
not repeat secret key queries for the same identity more than once. We show the security of the scheme via the following games. In each game, we define $X_i$ as the event that the adversary $\mathcal{A}$ wins in $\mathbf{Game_{i}}$.\\
$\mathbf{Game_{0}}$: This is the real security game. At the beginning of the game, $(\mathbf{A}, \mathbf{T_A})\leftarrow \mathsf{TrapGen}(1^n,1^m,q)$ is run and the adversary $\mathcal{A}$ is given $\mathbf{A}$. The challenger then samples $\beta\leftarrow \{0,1\}$ and keeps it secret. During the game, $\mathcal{A}$ may make random oracle queries, secret key queries, and the challenge query. These queries are handled as follows:\\
\indent$\bullet$\textsf{Hash queries}: When $\mathcal{A}$ makes a random oracle query to $\mathsf{H}$ on $id$, the challenger chooses a random matrix $\mathbf{U}_{id}\leftarrow \mathbb{Z}_q^n$ and locally stores the tuple $(id,\mathbf{U}_{id},\perp)$, and returns $\mathbf{U}_{id}$ to $\mathcal{A}$.\\
\indent$\bullet$\textsf{Secret key queries}: When the adversary $\mathcal{A}$ queries a secret key for $id$, the challenger uses the algorithm $\mathsf{SampleD}$ which takes $\mathbf{A},\mathbf{T}_A,\sigma,\mathbf{U}_{id}$ as input to compute $\mathbf{R}_{id}$  and returns $\mathbf{R}_{id}$ to $\mathcal{A}$.\\
\indent$\bullet$\textsf{Challenge ciphertext}: When the adversary $\mathcal{A}$ submits two messages $\sum_{\mathbf{m}_0}\alpha_{\mathbf{m}_0}|\mathbf{m}_0\rangle$ and $\sum_{\mathbf{m}_1}\alpha_{\mathbf{m}_1}|\mathbf{m}_1\rangle$ of equal length and a challenge identity $id^\ast$ with the restriction that $id^\ast$ is not equal to any identity requested in the previous phase. The challenger picks $\beta \overset{\$}{\leftarrow} \{0,1\}$, and encrypts $\sum_{\mathbf{m}_\beta}\alpha_{\mathbf{m}_\beta}|\mathbf{m}_\beta\rangle$ under $id^\ast$ by running the encryption algorithm $\mathsf{QEncrypt}$ to get $c^*=(|\psi\rangle,\mathbf{c_1})$, where $|\psi\rangle=|(\mathbf{x}+\lfloor \frac{q}{2}\rfloor\mathbf{m})~\textrm{mod}~q\rangle$ and $\mathbf{c_1}=(\mathbf{A}^\top \mathbf{s}+\mathbf{e})~\textrm{mod}~q$ and $\mathbf{x}=\mathbf{U}^\top \mathbf{s}+\mathbf{e_0}~\textrm{mod}~q$. It sends the ciphertext $c^*$ to the adversary $\mathcal{A}$.

\noindent At the end of the game, $\mathcal{A}$ outputs a guess $\beta^\prime$ for $\beta$. Finally, the challenger outputs $\beta^\prime$. By
definition, we have
\begin{equation}
\label{firstEquation}
|\Pr[X_0]-\frac{1}{2}|=|\Pr[\beta^\prime=\beta]-\frac{1}{2}|=\mathrm{Adv}_{\mathcal{A}}^{\mathrm{QIBE}}(\lambda).
\end{equation}
\noindent $\mathbf{Game_{1}}$: In this game, we change the way the random oracle queries to $\mathsf{H}$ are answered. When $\mathcal{A}$ queries the random oracle $\mathsf{H}$ on $id$, the challenger generates a pair $(\mathbf{U}_{id}, \mathbf{R}_{id})$ by first sampling
$\mathbf{r}_{id}^{i} \overset{\$}{\leftarrow} \mathcal{D}_{\mathbb{Z}^m,\sigma}$ to construct $\mathbf{R}_{id}$ and setting $\mathbf{U}_{id}= \mathbf{A}\cdot\mathbf{R}_{id}$. Then it locally stores the tuple $(id, \mathbf{U}_{id}, \perp)$, and returns $\mathbf{U}_{id}$ to $\mathcal{A}$. Here, we remark that when $\mathcal{A}$ makes a secret key query for $id$, the challenger uses the algorithm $\mathsf{SampleD}$ which takes $\mathbf{A},\mathbf{T}_A,\sigma,\mathbf{U}_{id}$ as input to compute $\mathbf{R}^{\prime}_{id}$  and returns $\mathbf{R}^{\prime}_{id}$ to $\mathcal{A}$. Note that $\mathbf{R}^{\prime}_{id}$ is independent from $\mathbf{R}_{id}$ that was generated in the simulation of the random oracle $\mathsf{H}$ on input $id$.  Due to Lemma \ref{fourLemma}, the distribution of $\mathbf{U}_{id}$ in $\mathbf{Game_{1}}$ is $n\cdot2^{-\Omega(n)}$-close to that of $\mathbf{Game_{0}}$ except for $2^{-\Omega(n)}$ fraction of $\mathbf{A}$ since we choose $\sigma > \sqrt{n + \log m}$. Therefore, we
have
\begin{equation}
|\Pr[X_1]-\Pr[X_0]|= n\cdot Q_{\mathsf{H}}\cdot 2^{-\Omega(n)}.
\end{equation}
\noindent$\mathbf{Game_{2}}$: In this game, we change the way secret key queries are answered. By the end of this
game, the challenger will no longer require the trapdoor $\mathbf{T}_{\mathbf{A}}$ to generate the secret keys. When $\mathcal{A}$
queries the random oracle on $id$, the challenger generates a pair $(\mathbf{U}_{id}, \mathbf{R}_{id})$ as in the previous game.
Then it locally stores the tuple $(id, \mathbf{U}_{id}, \mathbf{R}_{id})$ and returns $\mathbf{U}_{id}$ to $\mathcal{A}$. When $\mathcal{A}$ queries a secret key
for $id$, the challenger retrieves the unique tuple $(id, \mathbf{U}_{id}, \mathbf{R}_{id})$ from local storage and returns $\mathbf{R}_{id}$.
For any fixed $\mathbf{U}_{id}$, let $\mathbf{R}_{id,1}$ and $\mathbf{R}_{id,2}$ be random variables that are distributed according
to the distributions of $sk_{id}$ conditioning on $\mathsf{H}(id) = \mathbf{U}_{id}$ in $\mathbf{Game_{1}}$ and $\mathbf{Game_{2}}$, respectively.
Due to Lemma \ref{SixLemma}, we have $\Delta(\mathbf{r}_{id,1}^{i},\mathcal{D}_{\Lambda_{q}^{\mathbf{u}}(\mathbf{A}),\sigma})\leq 2^{-\Omega(n)}$ for $i=1,\cdots, n$. Due to Lemma \ref{fourLemma}, we have $\Delta(\mathbf{r}_{id,2}^{i},\mathcal{D}_{\Lambda_{q}^{\mathbf{u}}(\mathbf{A}),\sigma})\leq 2^{-\Omega(n)}$ for $i=1,\cdots, n$. Then we can get $\Delta(\mathbf{R}_{id,1},\mathbf{R}_{id,2})\leq n\cdot 2^{-\Omega(n)}$. Therefore we have
\begin{equation}
|\Pr[X_2]-\Pr[X_1]|= n\cdot Q_{\mathsf{ID}}\cdot 2^{-\Omega(n)}.
\end{equation}
\noindent$\mathbf{Game_{3}}$: In this game, we change the way the matrix $\mathbf{A}$ is generated. Concretely, the challenger
chooses $\mathbf{A}\overset{\$}{\leftarrow}\mathbb{Z}_{q}^{n\times m}$ without generating the associated trapdoor $\mathbf{T}_{\mathbf{A}}$. By Lemma \ref{SixLemma}, this makes only $2^{-\Omega(n)}$-statistical difference. Since the challenger can answer all the secret key queries without
the trapdoor due to the change we made in the previous game, the view of $\mathcal{A}$ is altered only by $2^{-\Omega(n)}$. Therefore, we have
\begin{equation}
\label{secondEquation}
|\Pr[X_3]-\Pr[X_2]|= 2^{-\Omega(n)}.
\end{equation}
\noindent$\mathbf{Game_{4}}$: In this game, we change the way the random oracle queries to $\mathsf{H}$ are answered and the challenge ciphertext is created. The challenger chooses an index $i^\ast\overset{\$}{\leftarrow}[Q_\mathsf{H}]$ and a matrix $\mathbf{U}\in\mathbb{Z}_q^{n\times n}$ uniformly at random.\\
\indent$\bullet$\textsf{Hash queries}: on $\mathcal{A}$'s $j$th distinct queries $id_j$ to $\mathsf{H}$, the challenger does the following: if $j=i^\ast$, then locally stores the tuple $(id_j, \mathbf{U}, \perp)$ and returns $\mathbf{U}$ to $\mathcal{A}$. Otherwise for $j\neq i^\ast$, the challenger selects $\mathbf{R}_{id_j}$ and computes $\mathbf{U}_{id_j}=\mathbf{A}\mathbf{R}_{id_j}$, then locally stores the tuple $(id_j, \mathbf{U}_{id_j}, \mathbf{R}_{id_j})$ and returns $\mathbf{U}_{id_j}$ to $\mathcal{A}$. \\
\indent$\bullet$\textsf{Challenge ciphertext}: when $\mathcal{A}$ produces a challenge identity $id^\ast$ (distinct from all its secret
key queries) and messages $\sum_{\mathbf{m}_0}\alpha_{\mathbf{m}_0}|\mathbf{m}_0\rangle$, $\sum_{\mathbf{m}_1}\alpha_{\mathbf{m}_1}|\mathbf{m}_1\rangle$, assume without loss of generality that $\mathcal{A}$ already queried $\mathsf{H}$ on $id^\ast$. If $id^\ast\neq id_{i^\ast}$, i.e., if the tuple $(id_{i^\ast}, \mathbf{U}, \perp)$  is not in local storage, then the challenger ignores the output of $\mathcal{A}$ and aborts the game (we denote this event as $\mathsf{abort}$). Otherwise, i.e., the $\mathsf{abort}$ does not happen(we denote this event as $\overline{\mathsf{abort}}$), the challenger picks $\beta \overset{\$}{\leftarrow} \{0,1\}$, and encrypts $\sum_{\mathbf{m}_\beta}\alpha_{\mathbf{m}_\beta}|\mathbf{m}_\beta\rangle$ under $id^\ast$ by running the encryption algorithm $\mathsf{QEncrypt}$ to get $c^*=(|\psi\rangle,\mathbf{c_1})$, where $|\psi\rangle=|(\mathbf{x}+\lfloor \frac{q}{2}\rfloor\mathbf{m})~\textrm{mod}~q\rangle$ and $\mathbf{c_1}=(\mathbf{A}^\top \mathbf{s}+\mathbf{e})~\textrm{mod}~q$, and $\mathbf{x}=\mathbf{U}^\top \mathbf{s}+\mathbf{e_0}~\textrm{mod}~q$. It sends the ciphertext $c^*$ to the adversary $\mathcal{A}$.\\
\indent Conditioned on the challenger not aborting, we claim that the view it provides to $\mathcal{A}$  in $\mathbf{Game_{4}}$ is statistically close to that in $\mathbf{Game_{3}}$. Therefore, we have
\begin{equation}
\label{midEquation}
\Pr[X_4\mid \overline{\mathsf{abort}}]=\Pr[X_3\mid \overline{\mathsf{abort}}].
\end{equation}
By a standard argument, the probability that the challenger does not abort during the simulation is $\frac{1}{Q_{\mathsf{H}}}$ (this is proved by considering a game in which the challenger can answer all secret key queries, so that the value of $i^\ast$ is perfectly hidden from $\mathcal{A}$). Therefore, we have
\begin{equation}
\Pr[\overline{\mathsf{abort}}]=\frac{1}{Q_{\mathsf{H}}}.
\end{equation}
\noindent$\mathbf{Game_{5}}$: In this game, we change the way the challenge ciphertext is created. \\
\indent$\bullet$\textsf{Challenge ciphertext}: when $\mathcal{A}$ produces a challenge identity $id^\ast$ (distinct from all its secret
key queries) and messages $\sum_{\mathbf{m}_0}\alpha_{\mathbf{m}_0}|\mathbf{m}_0\rangle$, $\sum_{\mathbf{m}_1}\alpha_{\mathbf{m}_1}|\mathbf{m}_1\rangle$, assume without loss of generality that $\mathcal{A}$ already queried $\mathsf{H}$ on $id^\ast$. If $id^\ast\neq id_i$, i.e., if the tuple $(id_i, \mathbf{U}, \perp)$  is not in local storage, then the challenger ignores the output of $\mathcal{A}$ and aborts the game (we denote this event as $\mathsf{abort}$). Otherwise, i.e., the $\mathsf{abort}$ does not happen (we denote this event as $\overline{\mathsf{abort}}$), the challenger picks $\beta \overset{\$}{\leftarrow} \{0,1\}$, and encrypts $\sum_{\mathbf{m}_\beta}\alpha_{\mathbf{m}_\beta}|\mathbf{m}_\beta\rangle$ under $id^\ast$ by using two random vector $\mathbf{b}^{\prime}\overset{\$}{\leftarrow}\mathbb{Z}_{q}^{n}, \mathbf{b}\overset{\$}{\leftarrow}\mathbb{Z}_{q}^{m}$ to get $c^*=(|\psi\rangle,\mathbf{c_1})$, where $|\psi\rangle=|(\mathbf{x}+\lfloor \frac{q}{2}\rfloor\mathbf{m})~\textrm{mod}~q\rangle$ and $\mathbf{c_1}=\mathbf{b}$, and $\mathbf{x}=\mathbf{b}^{\prime}$. It sends the ciphertext $c^*$ to the adversary $\mathcal{A}$. \\
\indent It can be seen that if $\left(\mathbf{A},\mathbf{U},\mathbf{c_1},\mathbf{x} \right)$ are valid LWE samples (i.e., $\mathbf{c_1}=(\mathbf{A}^\top \mathbf{s}+\mathbf{e})~\textrm{mod}~q$ and $\mathbf{x}=\mathbf{U}^\top \mathbf{s}+\mathbf{e_0}~\textrm{mod}~q$), the view of the adversary
corresponds to $\mathbf{Game_{4}}$. Otherwise (i.e., $\mathbf{c_1}\overset{\$}{\leftarrow}\mathbb{Z}_{q}^{m}, \mathbf{x}\overset{\$}{\leftarrow}\mathbb{Z}_{q}^{n}$), it corresponds to $\mathbf{Game_{5}}$. Therefore we have
\begin{equation}
\left|\Pr[X_5\wedge \overline{\mathsf{abort}}]-\Pr[X_4\wedge \overline{\mathsf{abort}}]\right| \leq \mathrm{Adv}_{\mathcal{B}}^{\mathrm{LWE}}(\lambda).
\end{equation}
Note that $\mathbf{c_1}, \mathbf{x}$  is statistically close to the uniform distribution over $\mathbb{Z}_{q}^{m}\times \mathbb{Z}_{q}^{n}$, so that
\begin{equation}
\label{lastEquation}
\Pr[X_5 \mid \overline{\mathsf{abort}}] = \frac{1}{2}.
\end{equation}
\indent According to equations  from (\ref{midEquation}) to (\ref{lastEquation}), we can get $$\left|\Pr[X_3 \mid \overline{\mathsf{abort}}]-\frac{1}{2}\right| \leq Q_{\mathsf{H}} \cdot \mathrm{Adv}_{\mathcal{B}}^{\mathrm{LWE}}(\lambda).$$
Then because $\overline{\mathsf{abort}}$ is independent of $X_3$, we can get
\begin{equation}
\label{additionalEquation}
\left|\Pr[X_3]-\frac{1}{2}\right| \leq Q_{\mathsf{H}} \cdot \mathrm{Adv}_{\mathcal{B}}^{\mathrm{LWE}}(\lambda).
\end{equation}
Finally, according to equations  from (\ref{firstEquation}) to (\ref{secondEquation}) together with equation (\ref{additionalEquation}),
we can get equation (\ref{finalEquation}).
\subsection{Security Network Protocols with Our QIBE}
A fundamental fact in quantum information theory is that unknown or random quantum states cannot be cloned \cite{1982A}. The quantum ciphertext state is $$ |\psi\rangle=\sum_\mathbf{m}\alpha_\mathbf{m}|(\mathbf{x}+\lfloor \frac{q}{2}\rfloor\mathbf{m})~\textrm{mod}~q\rangle.$$
For the adversary, the probability amplitude and the corresponding basis state of the ciphertext quantum states $|\psi\rangle$ is unknown, so they cannot try to copy it during its transmission. Then in the handshake protocol based on our QIBE, an attacker cannot copy and store the quantum ciphertext states of session keys whose confidentiality is protected by the secret identity key (which is called the long-term key). Thus, although the attacker gets the long-term key, they has no the quantum ciphertext of previous session keys to decrypt and cannot threat the security of the previous session key. In a word, all encrypted communications and sessions happened in the past cannot be retrieved.\par
 Therefore, the security protocol based on our QIBE has perfect forward security, which cannot achieve by the security protocol based on classic IBE.

\par

\section{Quantum Circuit realization}
\label{Quantum Circuit Realization of Quantum IBE}
\subsection{Quantum Circuit}
\label{Quantum Circuit1}
In order to analyze the realizability of our scheme in quantum circuit construction and estimate the quantum resources required by the scheme, in this section we give the specific quantum circuit implementation of our QIBE scheme.
The algorithms $\mathsf{QKeyGen}$ and $\mathsf{QExtract}$ of $\mathsf{QIBE}$ are classic algorithms and can be implemented with classic circuits. Thus we show the quantum circuit implementation of algorithms $\mathsf{QEncrypt}$ and $\mathsf{QDecrypt}$ of $\mathsf{QIBE}$ here. \par
\begin{figure}[h]
 \includegraphics[width=0.48\textwidth]{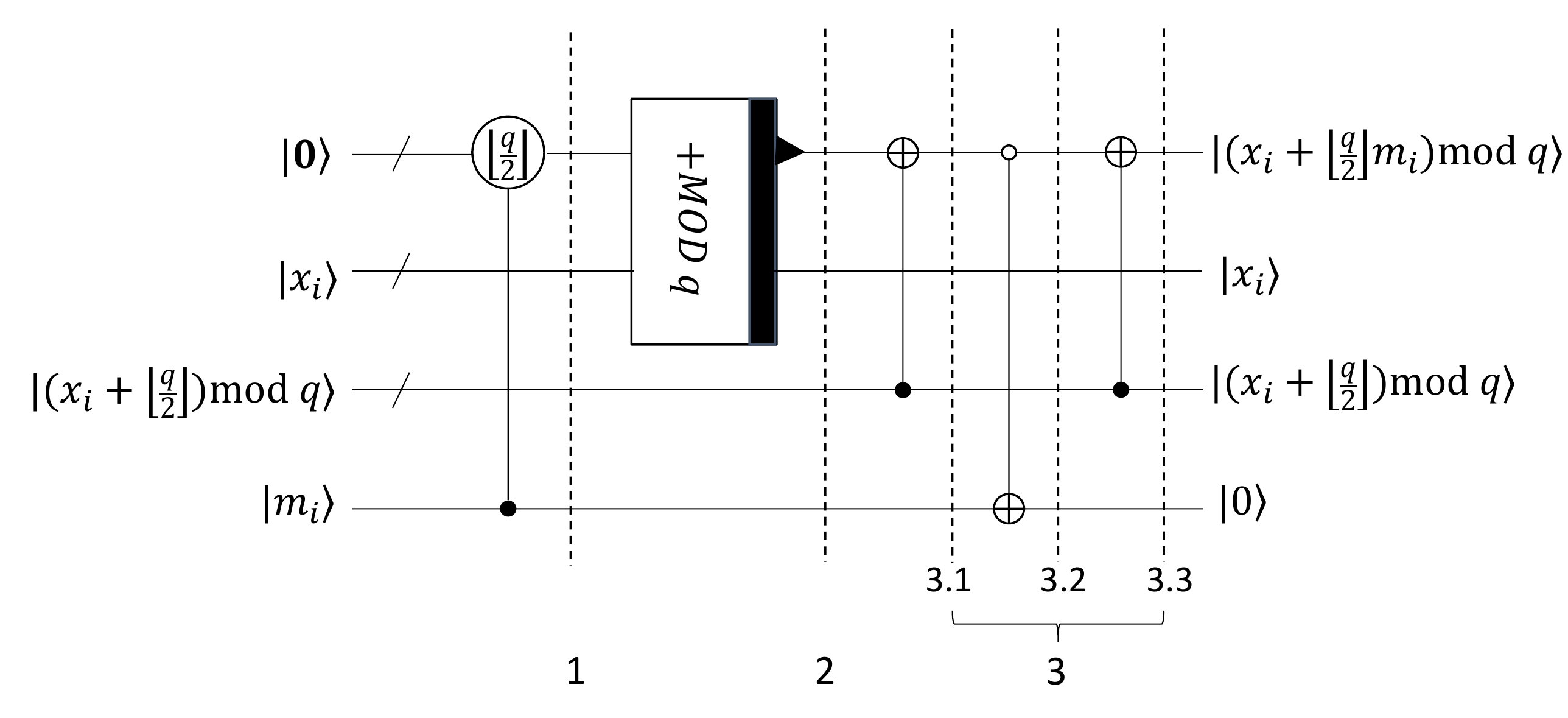}\\
 \caption{\small{The quantum circuit implementation of $\mathsf{QEncrypt}$}}
\label{quantumcircuitQencrypt}
 \centering
\end{figure}

\indent$\bullet$ Quantum circuit of the algorithm $\mathsf{QEncrypt}$: To simplify the description, we present the encryption quantum circuit of $|m_i\rangle$, and the encryption quantum circuit of $\sum_\mathbf{m}\alpha_\mathbf{m}|\mathbf{m}\rangle$ is its $n$-fold expansion. The quantum circuit implementation of the algorithm $\mathsf{QEncrypt}$ is shown in Figure \ref{quantumcircuitQencrypt}. In the first two steps of the encryption process, through the quantum controlled addition network and the quantum addition module $q$ network, we can get $|m_i\rangle|(x_i+\lfloor \frac{q}{2}\rfloor m_i)~\textrm{mod}~q\rangle$, and the third step is to get $|(x_i+\lfloor \frac{q}{2}\rfloor m_i)~\textrm{mod}~q\rangle$ from above result: \\
(\romannumeral1)~ The function of step 3.1 is to perform a bitwise exclusive OR of $|(x_i+\lfloor \frac{q}{2}\rfloor m_i)~\textrm{mod}~q\rangle$ and $|(x_i+\lfloor \frac{q}{2}\rfloor)~\textrm{mod}~q\rangle$. \\
(\romannumeral2)~ The effect of step 3.2 is to set $|m_i\rangle$ to $|0\rangle$. The specific analysis process is as bellow: If $|m_i\rangle$ is equal to $|1\rangle$, each bit of the result obtained by step 3.1 is $|0\rangle$, then the multi-control gate of step 3.2 makes $|m_i\rangle$ be set to $|(m_i+1)~\textrm{mod}~2\rangle=|0\rangle$. If $|m_i\rangle$ is equal to $|0\rangle$, each bit of the result obtained by step 3.1 is not all $|0\rangle$, then the multi-control gate of step 3.2 makes $|m_i\rangle$ be set to $|(m_i+0)~\textrm{mod}~2\rangle=|0\rangle$. \\
(\romannumeral3)~ Step 3.3 is the inverse of step 3.1, and its function is to offset the effect of step 3.1, that is, to recover $|(x_i+\lfloor \frac{q}{2}\rfloor m_i)~\textrm{mod}~q\rangle$ from the result obtained in step 3.1. \par
\indent$\bullet$ Quantum circuit of the algorithm $\mathsf{QDecrypt}$: The algorithm $\mathsf{QDecrypt}$ extracts $\sum_\mathbf{m}\alpha_\mathbf{m}|\mathbf{m}\rangle$ from the ciphertext quantum state. To simplify the description, we present the decryption quantum circuit of the ciphertext whose corresponding plaintext is $|m_i\rangle$, and the decryption quantum circuit of the ciphertext whose corresponding plaintext is $\sum_\mathbf{m}\alpha_\mathbf{m}|\mathbf{m}\rangle$ is its $n$-fold expansion. According to $\mathbf{Appendix}$, we can know that if $$\mathsf{abs}\left(\left(\left(x_i+\lfloor\frac{q}{2}\rfloor m_i\right)~\textrm{mod}~q-y_i\right)~\textrm{mod}~q
-\lfloor\frac{q}{2}\rfloor\right)<\lfloor\frac{q}{4}\rfloor,$$
 $|m_i\rangle = |1\rangle$; otherwise, $|m_i\rangle = |0\rangle$. Thus, before constructing the decryption quantum circuit, we need to construct the quantum absolute value circuit which implements the step 3 and step 5 of the $\mathsf{QDecrypt}$ process. \par
\begin{figure}[h]
 \includegraphics[width=0.48\textwidth]{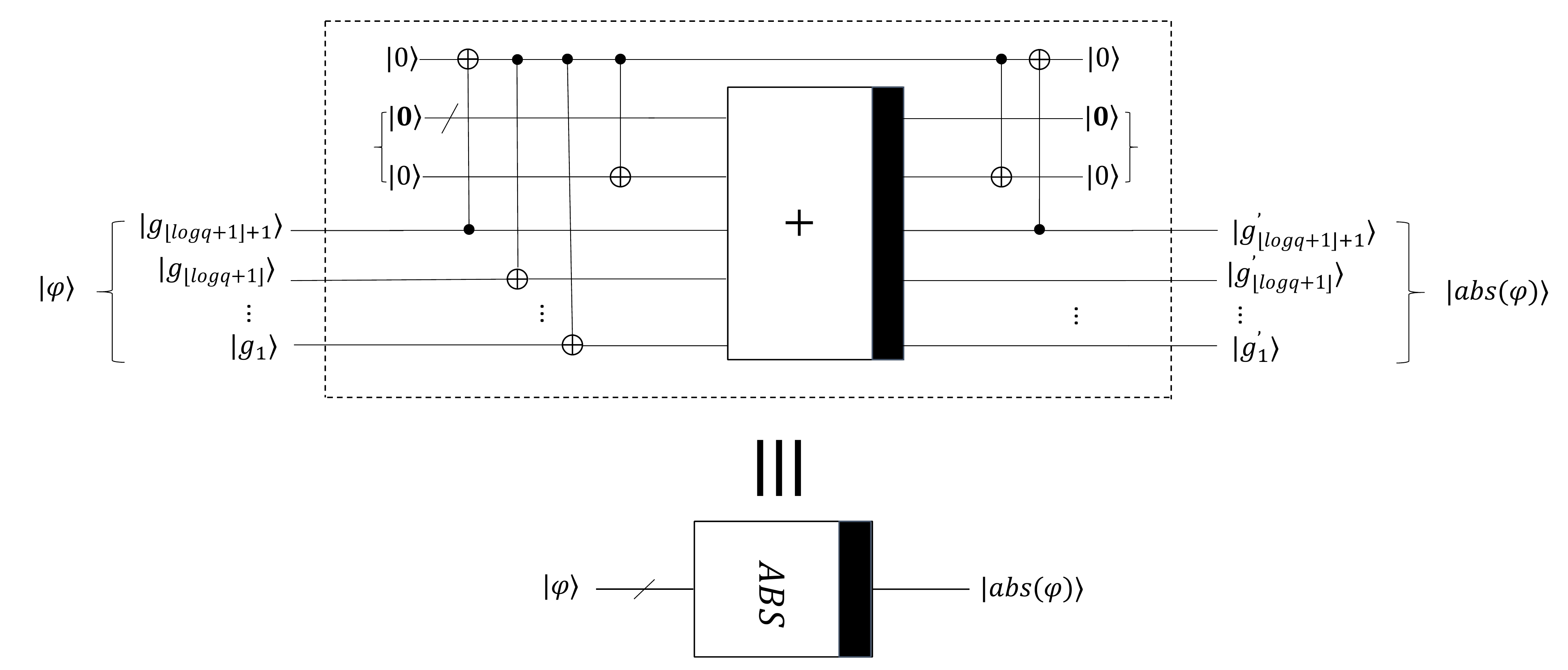}\\
 \caption{\small{The quantum circuit implementation of computing absolute value of $|\varphi\rangle$}}
\label{quantumcircuitabsolutevalue}
 \centering
\end{figure}

The analysis and construction process of the quantum absolute value circuit is as follows: Denote
$$|\varphi\rangle=\left|\left(\left(x_i+\lfloor\frac{q}{2}\rfloor m_i\right)~\textrm{mod}~q-y_i\right)~\textrm{mod}~q-\lfloor\frac{q}{2}\rfloor\right\rangle,$$
to prevent overflow when storing $|\varphi\rangle$, the size of the register to store $|\varphi\rangle$ should be $\lfloor \log q+1\rfloor+1$. Denote
$\left|g_j\right\rangle$ $\left(j=1,...,{\lfloor \log q+1\rfloor+1}\right)$ as the $j$-th bit of $\left|\varphi\right\rangle$ and $\left|g^\prime_j\right\rangle$ $\left(j=1,...,{\lfloor \log q+1\rfloor+1}\right)$ as the $j$-th bit of $\left|\mathsf{abs}\left(\varphi\right)\right\rangle$, where $\left|g_{\lfloor \log q+1\rfloor+1}\right\rangle$ and $\left|{g^\prime_{\lfloor \log q+1\rfloor+1}}\right\rangle$ are the most significant bits of $\left|\varphi\right\rangle$ and $\left|\mathsf{abs}\left(\varphi\right)\right\rangle$, respectively.
Denote ``$\overline{(\cdot)}$" is to reverse ``${(\cdot)}$'' bit by bit, for example $\left|\overline{101}\right\rangle=\left|010\right\rangle$. It is clear that if $|g_{\lfloor \log q+1\rfloor+1}\rangle=0$, $|\mathsf{abs}\left(\varphi\right)\rangle=|\varphi\rangle$; if $|g_{\lfloor \log q+1\rfloor+1}\rangle=1$, then $\left|\mathsf{abs}\left(\varphi\right)\right\rangle$ can be also expressed as bellow:
\[
\arraycolsep=1.4pt\def\arraystretch{1.8}
\begin{array}{ll}
&\left|\mathsf{abs}\left(\left(\left(x_i+\lfloor\frac{q}{2}\rfloor m_i\right)~\textrm{mod}~q-y_i\right)~\textrm{mod}~q-\lfloor\frac{q}{2}\rfloor\right)\right\rangle\\
&=\left|\overline{\left(\left(x_i+\lfloor\frac{q}{2}\rfloor m_i\right)\bmod q-y_i\right)\bmod q-\lfloor\frac{q}{2}\rfloor}+1\right\rangle\\
&=\left|2^{\lfloor \log q+1\rfloor}+\lfloor\frac{q}{2}\rfloor-\left(\left(x_i+\lfloor\frac{q}{2}\rfloor m_i\right)\bmod q-y_i\right)\bmod q\right\rangle,
\end{array}
\]
and we can also know that $|g_{\lfloor \log q+1\rfloor+1}\rangle=1$ in this situation from this result.
Thus, we can conclude that $|g_{\lfloor \log q+1\rfloor+1}\rangle=|{g^\prime_{\lfloor \log q+1\rfloor+1}}\rangle$.
For constructing the quantum absolute value circuit to calculate $|\mathsf{abs}\left(\varphi\right)\rangle$ from $|\varphi\rangle$, we use $|g_{\lfloor \log q+1\rfloor+1}\rangle$ as the control bit. If $\left|g_{\lfloor \log q+1\rfloor+1}\right\rangle=\left|1\right\rangle$, the circuit perform the operation of bitwise negation and adding 1 on the result; otherwise, no useful operations are performed on the input.
Then, we give the concrete quantum absolute value circuit and its simplified form shown in Figure \ref{quantumcircuitabsolutevalue}, which realizes $$\left|\varphi\right\rangle\rightarrow \left|\mathsf{abs}\left(\varphi\right)\right\rangle$$

To calculate the absolute value of $|\varphi\rangle$ which is $\lfloor \log q+1\rfloor+1$-qubits, a total of $2\lfloor \log q+1\rfloor+2$ Toffoli gates, $5\lfloor \log q+1\rfloor+9$ CNOT gates, and a total of $2\lfloor \log q+1\rfloor+2$ qubits are required for this quantum absolute value network.\par
\par
\begin{figure*}[htbp]
\centering
\includegraphics[scale=0.4]{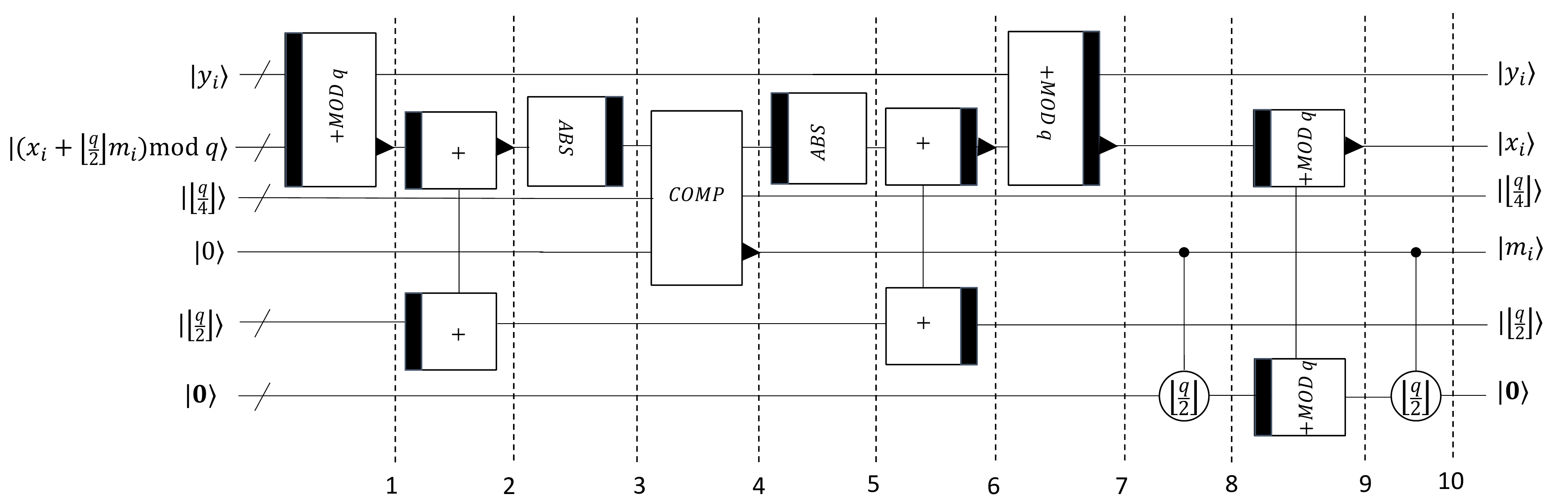}
\caption{\small{The quantum circuit implementation of $\mathsf{QDecrypt}$}}
\label{quantumcircuitQdecrypt}
\end{figure*}

Then, the quantum circuit implementation of the algorithm $\mathsf{QDecrypt}$ is shown in Figure \ref{quantumcircuitQdecrypt}, which marks the steps 1 to 10 in the process of the quantum decryption algorithm $\mathsf{QDecrypt}$ in Section \ref{Quantum Circuit1}.
%

%
\subsection{Quantum resource estimation}
 To measure the complexity of a quantum circuit, we should consider the number of quantum gates in the circuit and the total number of qubits used. In quantum circuits, it is meaningful to estimate the number of Hadamard gates, phase gate, CNOT gates and especially T gates. On the one hand, any unitary operator can be expressed exactly using single qubit and CNOT gates \cite{deutsch1985quantum}, and single qubit operation can be approximated to arbitrary accuracy using the Hadamard gate, phase gate and T gate \cite{nielsen2002quantum}. On the other hand, the structure of the fault-tolerant T gate is non-transverse and requires more complex and expensive technology to achieve it \cite{nielsen2002quantum,amy2013meet}. According to the trues that one $\lfloor\log q+1\rfloor$-controlled NOT gate can be decomposed into $2\lfloor\log q+1\rfloor-3$ Toffoli gates, and one Toffoli gate can be broken down into two Hadamard gates, one phase gate, seven T gates and six CNOT gates \cite{amy2013meet}, we estimate the quantum resources needed to encrypt the $n-$qubit quantum state $\sum_\mathbf{m}\alpha_\mathbf{m}|\mathbf{m}\rangle$ with the algorithm $\mathsf{QEncrypt}$ of QIBE, and decrypt the corresponding ciphertext with the algorithm $\mathsf{QDecrypt}$ of QIBE, including the numbers of Hadamard gate, phase gate, T gate, CNOT gate and the total qubits used, in which these gates constitute a universal quantum gate group. The quantum resources required by the quantum circuits of the encryption algorithm $\mathsf{QEncrypt}$ and the decryption algorithm $\mathsf{QDecrypt}$ are shown in the table \ref{quantumresource}. In order to save quantum resources, auxiliary bits can be reused according to the sequence of calculations in each circuit \cite{haner2020improved}. It can be seen from the table that the quantum resources required by our scheme increase linearly with the number of bits of the encrypted quantum plaintext.

\tikzset{global scale/.style={
    scale=#1,
    every node/.append style={scale=#1}
  }
}
\begin{table}[h]\small
\caption{Quantum resource.}
\label{quantumresource}
\begin{center}
\begin{tabular}{lccc}
\toprule  
Quantum resource& $\mathsf{QEncrypt}$     \\
\midrule  
Hadamard gate  & $2n\left(10\left(\lfloor \log q+1\rfloor\right)-3\right)$   \\
phase gate  & $n\left(10\left(\lfloor \log q+1\rfloor\right)-3\right)$   \\
T gate  & $7n\left(10\left(\lfloor \log q+1\rfloor\right)-3\right)$   \\
CNOT gate     &  $n\left(75.5\left(\lfloor \log q+1\rfloor\right)-12\right)$ \\
Qubit      &  $n\left(4\left(\lfloor \log q+1\rfloor\right)+4\right)$     \\
\midrule  
Quantum resource&  $\mathsf{QDecrypt}$ \\
\midrule  
Hadamard gate    &  $2n\left(34\left(\lfloor \log q+1\rfloor\right)+4\right)$   \\
phase gate  &  $n\left(34\left(\lfloor \log q+1\rfloor\right)+4\right)$   \\
T gate    &  $7n\left(34\left(\lfloor \log q+1\rfloor\right)+4\right)$   \\
CNOT gate      &  $n\left(269\left(\lfloor \log q+1\rfloor\right)+63\right)$\\
Qubit       & $n\left(6\left(\lfloor \log q+1\rfloor\right)+4\right)$  \\
\bottomrule 
\end{tabular}
\end{center}
\end{table}

\section{Conclusion}
\label{Conclusion}
In this paper, we proposed a kind of QIBE scheme based on the proposed classic IBE scheme \cite{DBLP:conf/stoc/GentryPV08}, and proved that it is fully secure. We construct the quantum circuit of $\mathsf{QEncrypt}$ of our scheme. Moreover, to implement the quantum circuit of $\mathsf{QDecrypt}$ of our scheme, we construct a quantum absolute value circuit and then give the quantum circuit of $\mathsf{QDecrypt}$ based on it. We estimate the quantum resources required for the quantum circuit of our scheme, including the numbers of Hadamard gate, phase gate, T gate, CNOT gate and total qubits used, and conclude that the quantum resources required by our scheme increase linearly with the number of bits of the encrypted quantum plaintext. Our QIBE scheme is suitable for quantum computing environment can encrypt both quantum messages and classic messages, and the classic KGC can still be used for the generation and distribution of identity secret key so that the cost can be reduced when the quantum scheme is implemented. In our QIBE scheme, the ciphertexts are transmitted in the form of a quantum state that is unknown to the adversary and cannot be copied and stored due to the no-cloning theorem. Thus, in the network security protocol based on our QIBE construction, even if the long-term key is threatened, the adversary cannot decrypt the previous ciphertexts to threat the previous session keys. Therefore, our QIBE scheme can make the network security protocol based on it have perfect forward security.\par
Our structure is one of the sixteen types of QIBE schemes described in section \ref{Definition of QIBE}, and the other fifteen types of QIBE schemes are yet to be studied. Moreover, the security of our scheme is based on the classic difficulty problem assumption the LWE assumption. Compared with the rapid development of quantum public-key encryption schemes based on the basic principles of quantum mechanics \cite{nikolopoulos2008applications,gao2009quantum,DBLP:journals/qip/WuY16,Chenmiao2017Qubit,yang2020quantum}, the design and research of the QIBE scheme based on the basic principles of quantum mechanics has a lot of room for development, which is also a very meaningful research direction.

\section{Acknowledgments}

This work was supported by National Natural Science Foundation of China (Grant No.
61672517), National Natural Science Foundation of China (Key Program, Grant No. 61732021).

\bibliographystyle{abbrv}
\bibliography{sample}

\newpage
\section*{Appendix A}
\label{Appendix}
In \cite{DBLP:conf/stoc/GentryPV08},  Gentry et al. proposed an identity based encryption $IBE=(\mathsf{KeyGen}, \mathsf{Extract},\mathsf{Encrypt}, \mathsf{Decrypt})$ from the learning with errors problem. In this scheme,
let integer parameters $n=\mathcal{O}(\lambda), m=\mathcal{O}(n),\sigma=\mathcal{O}(n^{0.5}), q=\mathcal{O}(m^{3.5})$, where $\lambda$ is a security parameter. \par
\indent$\bullet~\mathsf{KeyGen}:$ (1)Use the algorithm $\mathsf{TrapGen}(m,n,q)$ to select a uniformly random $n\times m-$matric $\mathbf{A}\in \mathbb{Z}_q^{n\times m}$ and $\mathbf{T_A}\in \mathbb{Z}_q^{m\times m}$ which is a good basis for $\Lambda_q^\perp(\mathbf{A})$. (2) Output the master key $\mathsf{mpk}=(\mathbf{A},q,m,n,\mathsf{H})$ and $\mathsf{msk}=(\mathbf{T_A})$. (3) In a word, $\mathsf{KeyGen}(\lambda, q,m,n)\rightarrow (\mathsf{mpk}=(\mathbf{A},q,m,n,\mathsf{H}),\mathsf{msk}=(\mathbf{T_A}))$. \par
\indent$\bullet~\mathsf{Extract}$: (1) Input $\mathsf{mpk}$, $\mathsf{msk}$ and an identity $id \in\{0,1\}^\ast$. (2) Use a hash function $\mathsf{H} : \{0, 1\}^\ast \rightarrow \mathbb{Z}_q^{n\times n}$ maps identities to $\mathbf{U}$, which is a $n\times n-$matric. (3) Take advantage of the algorithm $\mathsf{SampleD}$ to generate the secret key $sk_{id}=\mathbf{R}$ such that $\mathbf{r}^i=\mathsf{SampleD}(\mathbf{A},\mathbf{T_A}, \mathbf{u}^i,\sigma)$. It is easy to see that $\mathbf{U}=\mathbf{A}\mathbf{R}\bmod q$. (4) In a word, $\mathsf{Extract}(\mathsf{msk},\mathsf{mpk},id)\rightarrow sk_{id}=\mathbf{R}$.\par
\indent$\bullet~\mathsf{Encrypt}$: (1) To encrypt an $n-$bit message, take an identity $id$, $\mathsf{mpk}$ and the message $\mathbf{m}\in\{0,1\}^n$ as input, firstly choose a uniformly random $\mathbf{s}\leftarrow \mathbb{Z}_q^n$, $\mathbf{e}_0\in \mathcal{D}_{\mathbb{Z}^n,\sigma}$ and $\mathbf{e}\in \mathcal{D}_{\mathbb{Z}^m,\sigma}$. (2) Set $\mathbf{c}_0=\mathbf{U}^\top \mathbf{s}+\mathbf{e_0}+\lfloor \frac{q}{2}\rfloor\cdot\mathbf{m}~\textrm{mod}~q$ and $\mathbf{c_1}=(\mathbf{A}^\top \mathbf{s}+\mathbf{e})~\textrm{mod}~q$.
(3) In a word, $\mathsf{Encrypt}(id, \mathsf{mpk},\mathbf{m})\rightarrow c=(\mathbf{c_0},\mathbf{c_1})$.\par
\indent$\bullet~\mathsf{Decrypt}$: (1)Given the master public key $\mathsf{mpk}$, the private key $\mathbf{R}$, and the ciphertext $c=(\mathbf{c_0},\mathbf{c_1})$, compute $\mathbf{y}=\mathbf{R}^\top\mathbf{c_1}~\textrm{mod}~q$. (2)Then, compute $$\mathbf{b}=\left(\mathbf{c}_0-\mathbf{y}\right)~\textrm{mod}~q -\lfloor\frac{q}{2}\rfloor\cdot\mathbf{i}.$$ (3) Treat each coordinate of $\mathbf{b}=(b_1,\cdots,b_n)^\top$ as an integer in $\mathbb{Z}$, and set $m_i=1$ if $\mathsf{abs}(b_i)<\lfloor\frac{q}{4}\rfloor$, else $m_i=0$, where $i\in\{1,\cdots,n\}$. (4)Finally, return the plaintext $\mathbf{m}=(m_1,\cdots,m_n)^\top$. (5) In a word, $\mathsf{Decrypt}(id, \mathsf{mpk},sk_{id},c)\rightarrow \mathbf{m}$.\par

\vspace{3mm}
\noindent\textbf{Theorem 1.}
\label{theorem1}
Let integer parameters $n=\mathcal{O}(\lambda), m=\mathcal{O}(n),\sigma=\mathcal{O}(n^{0.5}), q=\mathcal{O}(m^{3.5})$. Consider a cipertext
$$(\mathbf{c_0},\mathbf{c_1})=\left(\mathbf{U}^\top \mathbf{s}+\mathbf{e_0}+\lfloor \frac{q}{2}\rfloor\cdot\mathbf{m},\mathbf{A}^\top \mathbf{s}+\mathbf{e}\right)~\textrm{mod}~q $$
of an $n$-bit message $\mathbf{m}$. Then the decryption algorithm $\mathsf{Decrypt}$ with the identity secret key $sk_{id}=\mathbf{R}$ can decrypt the ciphertext $c$ correctly with a probability $1-\mathsf{negl}(\lambda)$.

\end{document}